\documentclass[12pt]{article}

\usepackage[utf8]{inputenc}
\usepackage{amssymb,amsmath,amsfonts}
\usepackage{amsthm}
\usepackage[
  pdfusetitle,
  bookmarksnumbered,
  hidelinks,
  colorlinks,
  citecolor=blue,
  urlcolor=blue,
  linkcolor=blue,
  linktoc=page
]{hyperref}
\usepackage{dsfont}
\usepackage{graphicx}
\usepackage{caption}
\usepackage{subcaption}
\usepackage{pgfplots}
\usepackage{slashed}
\usepackage{cite}
\pgfplotsset{compat=1.15}

\graphicspath{{figures/}}

\usepackage{tikz}
\usetikzlibrary{shapes}

\def\Nm{{\mathcal{N}}}
\def\Om{{\mathcal{O}}}

\newcommand\wh\widehat
\newcommand{\vvev}[1]{\langle\!\langle #1 \rangle\!\rangle}
\newcommand{\vev}[1]{\langle #1 \rangle}

\newcommand\veps{{\varepsilon}}
\newcommand\zb{{\bar z}}

\newcommand\Dh{{\widehat{\Delta}}}
\newcommand\Dp{{\Delta_\phi}}
\newcommand\Dpp{{\Delta_{\phi^2}}}

\newcommand\gpp{\gamma_{\phi^2}}

\def\mathematica{{\texttt{Mathematica}}}

\DeclareMathOperator{\Disc}{Disc}

\DeclareMathOperator{\DR}{DR}

\usepackage{tikz}
\usetikzlibrary{shapes}
\usetikzlibrary{shapes.misc}

\tikzset{
  vtx/.style={
    circle,
    draw=blue,
    fill=blue,
    inner sep=1pt
  },
  wcirc/.style={
    circle,
    draw=white,
    fill=white,
    inner sep=2pt
  },
  bcirc/.style={
    circle,
    draw=black,
    fill=black,
    inner sep=1pt
  },
  dcirc/.style={
    circle,
    draw=blue,
    fill=blue,
    inner sep=1pt
  },
  rcirc/.style={
    circle,
    draw=red,
    fill=red,
    inner sep=1pt
  },
  phi/.style={
    thick
  },
  sigma/.style={
    thick,
    dashed
  },
  vl1/.style={
    thick,
    blue
  },
  vl2/.style={
    thick,
    dashed,
    blue
  },
  valign/.style={
    baseline={([yshift=-.55ex]current bounding box.center)}
  }
}

\title{\LARGE \bf Probing magnetic line defects \\ with two-point functions \\[0.5em]}
\author{
{\large Aleix Gimenez-Grau\footnote{gimenez@ihes.fr}} \\ ~ \\[0em]
{\normalsize \emph{Institut des Hautes Études Scientifiques, 91440 Bures-sur-Yvette, France}} \\[1em] }
\date{}
\begin{document}

\clearpage\maketitle
\thispagestyle{empty}

\begin{abstract}

~\\
\noindent{
This paper studies magnetic line defects in the Wilson-Fisher $O(N)$ model.
A powerful method to probe the system is to consider mixed two-point functions of the order parameter and the energy operator in the presence of the defect.
A recently developed dispersion relation allows us to bootstrap these mixed correlators to leading order in the $\veps$--expansion.
We also carry out explicit diagrammatic calculations, finding perfect agreement with the bootstrap, and we conclude extracting the new CFT data predicted by the two-point functions.
}

\end{abstract}

\newpage

\tableofcontents

\newpage

\section{Introduction}

Defects are extended operators in quantum field theory.
Although the most familiar type of defects are boundaries, there is a multitude of other examples, ranging from Wilson lines in gauge theory to impurities in statistical and condensed matter systems.
Defects often support new degrees of freedom, whose interactions with the bulk lead to very rich dynamics.
As usual in quantum field theory, the defect dynamics can be understood in terms of renormalization group flows that obey certain monotonicity properties, see \cite{Polchinski:2011im,Jensen:2015swa,Beccaria:2017rbe,Kobayashi:2018lil,Giombi:2020rmc,Wang:2021mdq,Cuomo:2021rkm,Beccaria:2021rmj,Rodriguez-Gomez:2022gif} for a partial list of references.
The present work focuses on defect conformal field theory (DCFT), in which the bulk and defect sit at their respective RG fixed-points.
The study of DCFT can be seen as a necessary first step towards a deeper understanding of defect RG flows, but it also has important applications on its own, for example to impurities in critical condensed matter systems.

We concentrate on the critical $O(N)$ model, arguably one of the most well-understood CFTs in $d\ge2$, which also has extensive experimental applications \cite{Pelissetto:2000ek}.
In particular, we turn on a background magnetic field localized along a one-dimensional line, which in field-theory language is described by\footnote{The $O(N)$ symmetry allows us to orient the background field in the direction $\phi_1$.}
\begin{align}
  S = S_{O(N)} -h_0 \int d\tau \, \phi_1(\tau, x_\bot=0) \, .
  \label{eq:def-defect}
\end{align}
This action defines a defect known in the condensed-matter literature as the pinning-field defect \cite{Allais:2014fqa,2014arXiv1412:3449A,2017PhRvB:95a4401P}, but here we follow  \cite{Cuomo:2021kfm} and call it the localized magnetic field line defect.
The defect \eqref{eq:def-defect} breaks the bulk symmetry down to $O(N-1)$, while for $N\,{=}\,1$ the $\mathbb{Z}_2$ symmetry is completely broken.
This makes the localized magnetic field defect the one-dimensional analog of the normal boundary CFT, see \cite{Metlitski:2020cqy,Padayasi:2021sik,Toldin:2021kun} for a recent discussion.

In $d < 4$ the scalar $\phi_1$ is relevant and triggers an RG flow on the defect.
By tuning the bulk couplings to the critical point, one reaches the defect CFT that we study in this work (note that $h_0$ does not need tuning \cite{Cuomo:2021kfm}).
At the experimental level, defect CFTs are characterized by a new set of critical exponents.
Although we are not aware of an experimental determination of critical exponents for the localized magnetic field defect, it is likely the system can be analyzed in the lab.
For example, in the realization of the $O(3)$ model as a quantum critical antiferromagnet, the defect \eqref{eq:def-defect} should correspond to turning on a localized staggered magnetic field, and then measuring the effects at large distances and times.

Because experiments might be able to measure the critical exponents of the localized magnetic field defect, it is an important challenge to predict them.
To date, two methods have been used to determine the critical exponents: perturbation theory and lattice simulations.
On the perturbative side, the nice paper \cite{Cuomo:2021kfm} performed $4-\veps$ and large-$N$ calculations, but see also \cite{Allais:2014fqa,Rodriguez-Gomez:2022gbz,Gimenez-Grau:2022czc,Giombi:2022vnz} for previous as well as related work.
From the numerical side, one can model \eqref{eq:def-defect} with classical or quantum lattice simulations.
For the Ising universality class \cite{2014arXiv1412:3449A,2017PhRvB:95a4401P}, one may consider the classical Hamiltonian $H = -K \sum_{\langle ij\rangle} \sigma_i \sigma_j - h_0 \sum_{i\in\text{line}} \sigma_i$, where the last term corresponds to the line defect.
On the other hand, for the $O(3)$ model one can use a bilayer quantum Heisenberg Hamiltonian with an extra term $H_{\text{defect}} = h_0 S_{i_0}^z$, where $i_0$ specifies the lattice site where the defect is located \cite{2017PhRvB:95a4401P}.
Overall, the perturbative and numerical results are in good agreement \cite{2017PhRvB:95a4401P,Cuomo:2021kfm}.

In parallel to these developments, it has become clear in the last decade that the conformal bootstrap provides a competitive alternative to perturbative calculations and Monte Carlo methods \cite{Poland:2018epd}.
Broadly speaking, the bootstrap wants to understand the space of CFTs and solve specific models relying on basic principles like symmetry and unitarity.
The bootstrap philosophy also applies to defects \cite{Liendo:2012hy,Billo:2016cpy}, and significant progress has been made recently.\footnote{The literature is too vast to review here, but see \cite{Gaiotto:2013nva,Billo:2016cpy,Lauria:2017wav,Lauria:2018klo,Isachenkov:2018pef,Gliozzi:2015qsa,Bissi:2018mcq} for a partial list of references.}
The defect bootstrap is interested in answering two questions.
First, what is the space of consistent conformal defects that a given bulk theory can support?
And second, given a defect CFT of interest, how to determine critical exponents and other observables?
In this work, we give answers to both of these questions for the localized magnetic line defect in the $\veps$--expansion.
Ultimately, the goal is to bootstrap conformal defects in a non-perturbative fashion, to provide an alternative to lattice and diagrammatic methods.
As we discuss below, the present work provides an important step towards this goal.

\subsection*{Summary of results}

It is good to summarize the main results of the paper before getting into the details of their derivation.
Section \ref{sec:bootstrap} bootstraps the mixed two-point functions of the order parameter $\phi_a$ and energy operator $\phi^2$ in the presence of a conformal line defect.
Recent techniques \cite{Barrat:2022psm,Bianchi:2022ppi,Lemos:2017vnx} based on the Lorentzian inversion formula \cite{Caron-Huot:2017vep} and the conformal dispersion relation \cite{Carmi:2019cub} allow us to determine the two-point functions to order $O(\veps)$ in the $\veps$--expansion.
In section \ref{sec:feyndiag}, the bootstrap results are compared against diagrammatic calculations using the Lagrangian description \eqref{eq:def-defect}, and we find perfect agreement.
Due to the symmetries broken by the defect, the two-point functions are not kinematically fixed, so they contain dynamical information that can be extracted with operator product expansions, as we do in detail in section \ref{sec:ope-analysis}.
For the reader's convenience, we made sections \ref{sec:bootstrap}, \ref{sec:feyndiag} and \ref{sec:ope-analysis} as self-contained as possible, so that they can been read independently, and we summarized ideas for future research in the concluding section \ref{sec:conclusions}.

Let us comment more on the bootstrap calculation, focusing for concreteness on the Ising universality class ($N=1$).
The bootstrap does not rely on the Lagrangian description, and instead it uses: \emph{i}) scaling dimensions and OPE coefficients for the Wilson-Fisher (WF) fixed point; \emph{ii}) the bulk operator $\phi^3$ is a conformal descendant in the WF fixed point \cite{Rychkov:2015naa}; \emph{iii}) due to broken translations, the defect supports a displacement operator; \emph{iv}) correlators with the displacement satisfy a Ward identity \cite{Billo:2016cpy}.
Principles \emph{i-iv}) apply to any conformal defect in the Wilson-Fisher fixed point.
To narrow down our search, we also impose that \emph{v}) the defect breaks the global $\mathbb{Z}_2$ symmetry; \emph{vi}) the defect Regge trajectories are analytic down to spin zero.
Provided \emph{i-vi}) hold, the bootstrap determines uniquely the two-point functions, with no free parameters left.
We stress that using mixed-correlator bootstrap is essential to fix all the free coefficients.

As mentioned above, one goal of the bootstrap is to understand which conformal defects can be supported in a given bulk theory.
Our results give a sharp answer for $\mathbb Z_2$-breaking line defects in the Wilson-Fisher model: if condition \emph{vi}) holds, then there is a unique defect, which coincides with the localized magnetic field defect \eqref{eq:def-defect}.
Recently, there has been significant progress on classifying defects for free bulk fields \cite{Lauria:2020emq,Behan:2020nsf,Behan:2021tcn,Herzog:2022jqv}.
Our work and \cite{Gimenez-Grau:2021wiv,Barrat:2021yvp} may be seen as the first steps towards generalizing these results for weakly-coupled models.\footnote{There has also been similar progress for BCFT, see for example \cite{Bissi:2018mcq,Dey:2020jlc,Gimenez-Grau:2020jvf}, although the methods are somewhat different than the ones here.}
Alternatively, our results are also an extension of the analytic mixed-correlator bootstrap for the Wilson-Fisher fixed point \cite{Bertucci:2022ptt}, to a case where defects are present.
Although we use perturbation theory, it is plausible that conditions \emph{i-vi}) also determine uniquely the localized magnetic line defect in the (non-perturbative) critical 3d Ising model.
This information should be helpful for future non-perturbative bootstrap studies, as we discuss more fully in the conclusions.

We should also mention that for technical reasons discussed later, we only considered single correlator bootstrap for $N>1$.
In this case the defect breaks global symmetry $O(N) \to O(N-1)$, and there exists a tilt operator due to this breaking \cite{Cuomo:2021kfm,Padayasi:2021sik}.
For $N>1$ we can fix the two-point function up to one free coefficient, but we believe that if we bootstrapped the mixed correlators, the final result would be uniquely determined.
More generally, our results make it plausible that in the $O(N)$ model there is a unique line defect that breaks the symmetry to $O(N-1)$ (see \cite{Cuomo:2022xgw} for a line defect that preserves the global symmetry).
It would be fascinating to find more evidence for this hypothesis using non-perturbative bootstrap methods.

Besides conformal bootstrap, we also study the localized magnetic line defect using diagrammatic perturbation theory.
We start with bulk one-point functions and bulk-defect two-point functions, which are are kinematically fixed but contain dynamical information in their normalization, often called CFT data.
We then compute bulk-bulk two-point functions, which are not kinematically fixed and capture at once CFT data for infinite families of operators.
When comparison is possible, we find perfect agreement between the bootstrap and diagrammatic results.
The main challenge in the diagrammatic approach is to evaluate the Feynman integrals, but after these integrals have been determined, they can be reused in multiple situations.
For example, there is little difference between the diagrammatic calculation for the line defect in the $O(N)$ model for $N=1$ or $N>1$, while the bootstrap for $N>1$ is significantly harder.
Similarly, we can access the two-point function of the energy operator $\vvev{ \phi^2 \phi^2}$, which is also difficult to obtain from bootstrap.
We hope the diagrams computed here might be of use in the future, for instance to study other line defects such as \cite{Cuomo:2022xgw,Popov:2022nfq,Giombi:2022vnz}.
Finally, we extract the CFT data predicted by the two-point functions by means of conformal block decompositions.
Much of this CFT data was previously unknown, and we hope it might be of interest in future studies of the localized magnetic field line defect.

\paragraph{Note added:} When this work was being finalized, we became aware of \cite{LDE}, which partially overlaps with our work. We coordinated with the authors for a simultaneous submission.

\section{Analytic bootstrap calculation}
\label{sec:bootstrap}

In this section, we study the localized magnetic line defect with analytic bootstrap techniques \cite{Caron-Huot:2017vep,Alday:2017zzv} adapted to conformal defects \cite{Lemos:2017vnx,Barrat:2022psm,Bianchi:2022ppi}.
The defect bootstrap has been discussed extensively in the literature, but to keep the paper self contained, we provide a brief review in appendix \ref{sec:review-dcft}.
For the sake of clarity, we focus first on the Ising universality class $N=1$, and we comment on $N>1$ in section \ref{sec:on-model}.
Because we are interested in defects that break the $\mathbb Z_2$ symmetry, we can consider the two-point functions of the order parameter $\phi$ and the energy operator $\phi^2$:
\begin{align}
 \vvev{ \phi(x_1) \phi(x_2) }
 = \frac{F_{\phi\phi}(r, w)}{|\vec x_1|^\Dp |\vec x_2|^\Dp} \, , \qquad
 \vvev{ \phi(x_1) \phi^2(x_2) }
 = \frac{F_{\phi\phi^2}(r, w)}
        {|\vec x_1|^\Dp |\vec x_2|^{\Delta_{\phi^2}}} \, .
 \label{eq:corrs-to-boot}
\end{align}
Here and below the double-bracket notation $\vvev{ \Om \ldots }$ indicates that an expectation value is taken in the presence of a defect.
Furthermore, we split spacetime vectors $x = (\hat x, \vec x)$ into parallel coordinates $\hat x$ and orthogonal coordinates $\vec x$ with respect to the defect.
The correlators $F_{\phi\phi}$ and $F_{\phi\phi^2}$ depend non-trivially on the cross-ratios $r$, $w$ defined in \eqref{eq:cross-ratios-rw}.
By means of operator product expansions, these correlators contain information about the $\mathbb{Z}_2$-even and $\mathbb{Z}_2$-odd sectors of the model.
Below we bootstrap these correlators analytically to leading order in the $\veps$--expansion.
Initially, the bootstrap result depends on four undetermined constants
\begin{align}
 a_\phi = a_\phi^{(0)} + \veps \, a_\phi^{(1)} + O(\veps^2) \, , \qquad
 a_{\phi^2} = a_{\phi^2}^{(0)} + \veps \, a_{\phi^2}^{(1)} + O(\veps^2) \, ,
 \label{eq:free-params}
\end{align}
where $a_\Om \propto \vvev{\Om(x)}$ are normalizations of one-point functions, see \eqref{eq:one-pt-fun}.
However, the correlators $F_{\phi\phi}$ and $F_{\phi\phi^2}$ obey several consistency conditions that allow to fix the free parameters uniquely, as explained in section \ref{sec:free-coeffs}.

\subsection{Structure of the calculation}
\label{sec:struct-calc}

Our goal is to use analytic bootstrap to determine the correlators \eqref{eq:corrs-to-boot}, and because the calculation that follows is somewhat involved, we start with a review of the most important ideas.

\paragraph{Dispersion relation}
An important technical tool in our calculation is the defect dispersion relation \cite{Barrat:2022psm,Bianchi:2022ppi}, which is defined by the integral transform
\begin{align}
 \DR \big[ F(r,w) \big]
 \equiv \int_0^r \frac{dw'}{2 \pi i}
 \frac{w (1 - w') (1 + w')}{ w' (w'-w) (1 - w w')}
 \Disc F(r,w') \, ,
 \label{eq:disp-rel}
\end{align}
and the discontinuity is $\Disc F(r,w) = F(r,w+i0) - F(r,w-i0)$.
As shown in \cite{Barrat:2022psm,Bianchi:2022ppi}, this integral reconstructs the correlator from its single discontinuity $F(r,w) = \DR \big[ F(r,w) \big]$, up to a caveat discussed at the end of the section.
The advantage of the dispersion relation is that the discontinuity is a much simpler object than $F(r,w)$, specially in perturbation theory.
Once the discontinuity is known, reconstructing the correlator with equation \eqref{eq:disp-rel} becomes a mechanical, albeit sometimes challenging, calculation.
Let us see how this works for $F_{\phi\phi}$ and $F_{\phi\phi^2}$.

\paragraph{Bulk operator families:}
To compute the discontinuities, we need a preliminary discussion of the spectrum of the Wilson-Fisher model, while section \ref{sec:review-WF} will provide a more detailed treatment.
In the Wilson-Fisher model the fundamental scalar $\phi$ is odd under $\mathbb{Z}_2$ symmetry.
The remaining primary operators are composites of $\phi$ that organize in families with even/odd $\mathbb{Z}_2$ charge.
Loosely speaking, these operators take the form\footnote{This is schematic because two derivatives $\partial^2$ and two fields $\phi^2$ have the same tree-level quantum numbers. There are also inequivalent ways to act with the derivatives that lead to different conformal primaries.}
\begin{align}
 \Om^+_{\ell,n}
 \sim (\partial^2)^n
      \partial_{\mu_1} \ldots \partial_{\mu_\ell} \phi^2 \, , \qquad
 \Om^-_{\ell,n}
 \sim (\partial^2)^n
      \partial_{\mu_1} \ldots \partial_{\mu_\ell} \phi^3 \, .
  \label{eq:defOpOm}
\end{align}
The scaling dimension of these operators follows from the fundamental field $\Dp = \frac{d-2}{2} + O(\veps^2)$, up to small corrections in $\veps$.
For example, the $\mathbb{Z}_2$-even operators have dimension
\begin{align}
 \Delta_{\Om^+_{\ell,n}}
 = 2\Delta_\phi + \ell + 2n + \gamma_{\Om^+_{\ell,n}} \, , \qquad
 \gamma_{\Om^+_{\ell,n}}
 = \veps \gamma^{(1)}_{\Om^+_{\ell,n}}
 + \veps^2 \gamma^{(2)}_{\Om^+_{\ell,n}}
 + \ldots \, .
\end{align}
The twist is defined as $\Delta-\ell$, so $n=0$ is the leading-twist trajectory, while $n\ge1$ have higher twist.
Except for the leading $\mathbb{Z}_2$-even trajectory, in general at each $n,\ell$ there exist multiple inequivalent primaries of the form \eqref{eq:defOpOm} with the same tree-level dimensions.
Quantum corrections lift this degeneracy, but because we work in perturbation theory, we often encounter complications due to mixing between nearly-degenerate operators.

Much is known about the $\mathbb{Z}_2$-even operators $\Om_{\ell,n}^+$, and this information plays an important role for the bootstrap.
The scalar $\phi^2 = \Om^+_{0,0}$ receives anomalous dimension at leading order, and has order-one OPE coefficient:
\begin{align}
 & \Delta_{\phi^2}
 = 2 \Delta_\phi + \frac{\veps}{3} + O(\veps^2) \, , \qquad
 \lambda_{\phi\phi\phi^2}^2
 = \big( \lambda_{\phi\phi\phi^2}^{(0)} \big)^2
 + \big( \lambda_{\phi\phi\phi^2}^{(1)} \big)^2 \veps
 + \ldots
\end{align}
The remaining operators at $n=0$ are higher-spin currents in the free theory, and at the order we work, they have tree-level dimensions and order-one OPE coefficients:
\begin{align}
 & \Delta_{\Om^+_{\ell,0}}
 = 2 \Dp + \ell + O(\veps^2) \, , \qquad
 \lambda_{\phi\phi\Om^+_{\ell,0}}^2
 = \big( \lambda_{\phi\phi\Om^+_{\ell,0}}^{(0)} \big)^2
 + \big( \lambda_{\phi\phi\Om^+_{\ell,0}}^{(1)} \big)^2 \veps
 + \ldots
 \label{eq:OpDim}
\end{align}
Finally, the subleading trajectories $n \ge 1$ have anomalous dimensions at leading order \cite{Kehrein:1992fn}
\begin{align}
  \Delta_{\Om^+_{\ell,n}}
  = 2\Dp + \ell+2n
  + \gamma^{(1)}_{\Om^+_{\ell,n}} \veps
  + O(\veps^2) \, , \qquad
  n \ge 1 \, ,
\end{align}
and as observed in \cite{Alday:2017zzv,Henriksson:2018myn}, the OPE coefficients take the form
\begin{align}
  \lambda_{\phi\phi \Om^+_{\ell,1}}^2
  = O(\veps^2) \, , \qquad
  \lambda_{\phi\phi \Om^+_{\ell,n\ge2}}^2
  = O(\veps^3) \, .
  \label{eq:orders-ng1}
\end{align}

\paragraph{Correlator $F_{\phi\phi}$:}
Keeping this information in mind, we come back to the study of line defects in the $\veps$--expansion.
As reviewed in appendix \ref{sec:block-exp-2pt}, two-point functions admit a bulk block expansion \eqref{eq:blk-block-exp}.
For $F_{\phi\phi}$, the block expansion is in terms of $\mathbb{Z}_2$--even operators $\Om_{\ell,n}^+$, and fortunately several simplifications occur at order $O(\veps)$.
On the one hand, the $n=1$ family appears with tree-level dimension because it is multiplied by an $O(\veps)$ OPE coefficient, while trajectories with $n\ge 2$ do not appear because of \eqref{eq:orders-ng1}.
On the other hand, the leading-twist trajectory also has tree-level dimensions \eqref{eq:OpDim}.
All in all, the expansion has the form
\begin{align}
  \xi^{\Delta_\phi} F_{\phi\phi}(r,w)
& = 1
    + \lambda_{\phi\phi\phi^2} a_{\phi^2} f_{\Delta_{\phi^2},0} 
    + \sum_{\substack{\ell=2 \\ \ell \text{ even}}}^\infty 2^{-\ell}
    \lambda_{\phi\phi\Om_{\ell,0}^+}
    a_{\Om_{\ell,0}^+}
    f_{2\Delta_\phi+\ell,\ell} 
    \notag \\
& \quad
    + \sum_{\substack{\ell=0 \\ \ell \text{ even}}}^\infty 2^{-\ell}
    \lambda_{\phi\phi\Om_{\ell,1}^+}
    a_{\Om_{\ell,1}^+}
    f_{2\Delta_\phi+\ell+2,\ell} 
    + O(\veps^2) \, .
 \label{eq:PP-ising-block}
\end{align}
The coefficients $a_\Om$ are the normalization of the one-point function of $\Om$ in the presence of the defect, see \eqref{eq:one-pt-fun}.
We do not know these one-point functions a priori: they will be predictions from our calculation.
The conformal blocks $f_{\Delta,\ell}$ depend on the cross-ratios $r,w$, although we drop the dependence for clarity.
Finally, the cross-ratio $\xi$ is defined in terms of $r,w$ in \eqref{eq:cr-xi}, while the $2^{-\ell}$ ensure consistency with our normalization conventions.

The crucial observation is that the discontinuity kills all operators with tree-level dimension in the block decomposition \eqref{eq:PP-ising-block}.
Indeed, it is not hard to show that the discontinuity of a general block reads
\begin{align}
 \Disc \big[ \xi^{-\frac12(\Delta_1+\Delta_2)}
 f^{\Delta_{12}}_{\Delta_1+\Delta_2+\ell+2n+\gamma,\ell}(r,w) \big]
 \approx \gamma \Disc
   \big[ \xi^{-\frac12(\Delta_1+\Delta_2)}
 \partial_\Delta f^{\Delta_{12}}_{\Delta_1+\Delta_2+\ell+2n,\ell}(r,w) \big]
   \, ,
  \label{eq:disc-block}
\end{align}
provided the anomalous dimension $\gamma$ is parametrically small.
The important message is that the right-hand side is proportional to the anomalous dimension $\gamma$, so operators with tree-level dimension are killed.
A caveat is that this argument does not work for operators below the double-twist dimension $\Delta<\Delta_1+\Delta_2$, so for example the discontinuity of the identity operator $\Delta =0$ has to be computed independently.

The upshot is that acting with $\Disc$ on the block expansion \eqref{eq:PP-ising-block} leads to dramatic simplifications at order $O(\veps)$:
\begin{align}
\label{eq:sing-part-wf}
 \Disc F_{\phi\phi}(r,w) \big|_{O(\veps)}
 & = \Disc \xi^{-\Delta_\phi}
   + \frac{\veps}{3} \,
     \lambda^{(0)}_{\phi\phi\phi^2} \, a^{(0)}_{\phi^2}
     \Disc \left[ \xi^{-1} \partial_\Delta f_{2, 0} \right] \, .
\end{align}
The first term is the contribution from the identity block, while the second comes from the anomalous dimension of $\phi^2$.
Since we work at order $O(\veps)$ and the second term is multiplied by $\veps$, we replaced by the tree-level OPE coefficients $\lambda^{(0)}_{\phi\phi\phi^2} \, a^{(0)}_{\phi^2}$ as well as the tree level dimensions $\Dpp = 2\Dp = 2$.
The leading-twist family dropped out because the anomalous dimensions are higher order, see \eqref{eq:OpDim}, while the first subleading family dropped out because although the anomalous dimensions are order $O(\veps)$, the OPE coefficients are also $O(\veps)$.

To summarize, we started with the correlation function $F_{\phi\phi}$, of which we know almost nothing except the structure of the block expansion \eqref{eq:PP-ising-block}.
By instead considering its discontinuity, we obtained a dramatically simpler expression \eqref{eq:sing-part-wf}, where the only unknown parameter is $a_{\phi^2}^{(0)}$.
We can then apply the machinery of the dispersion relation \eqref{eq:disp-rel} to reconstruct the correlator.

\paragraph{Correlator $F_{\phi\phi^2}$:}
A completely analogous discussion goes through for $F_{\phi\phi^2}$.
In this case, the bulk OPE has the schematic form $\phi \times \phi^2 \sim \phi + \sum_{\ell,n} \Om_{\ell,n}^-$.
Because $\phi$ sits below the triple-twist dimension $\Dp+\Dpp$ equation \eqref{eq:disc-block} does not apply and we compute its contribution separately.
On the other hand, the higher-twist trajectories $\Om_{\ell,n\ge1}^-$ do not contribute because their OPE coefficients are order $O(\veps)$.
Finally, care is needed with the triple-twist family $\Om_{\ell,0}^-$, because it has $O(\veps)$ anomalous dimensions and order-one coefficients \cite{Kehrein:1992fn,Bertucci:2022ptt}:\footnote{Here there is a subtlety because this family is degenerate. However, the results of \cite{Bertucci:2022ptt} show that we can treat it as non-degenerate at the order we work. We discuss this fully in section \ref{sec:review-WF}. }
\begin{align}
 \Delta_{\Om^-_{\ell,0}}
&= \Dp
 + \Delta_{\phi^2}
 + \frac23 \frac{(-1)^\ell}{\ell+1} \, \veps
 + O(\veps^2) \, , \qquad
 \lambda^2_{\phi\phi^2\Om^-_{\ell,0}}
 = O(1) \, .
\end{align}
As a result, the discontinuity reduces to
\begin{align}
 \Disc F_{\phi\phi^2}(r,w) \big|_{O(\veps)}
 & = \lambda_{\phi\phi^2\phi} a_{\phi}
   \Disc \left[ \xi^{-(\Dp+\Dpp)/2} f^{\Dp-\Delta_{\phi^2}}_{\Dp, 0} \right]
   \notag \\
 & \quad
   + \frac{2\veps}{3}
     \sum_{\substack{\ell=2\\ \ell \text{ even}}}^\infty
     \frac{2^{-\ell}}{\ell + 1} \,
     \lambda^{(0)}_{\phi\phi^2\Om^-_{\ell,0}} a^{(0)}_{\Om^-_{\ell,0}}
     \Disc \left[ \xi^{-3/2} \, \partial_\Delta f^{-1}_{3+\ell, \ell} \right] \, .
\label{eq:sing-part-phi-phi2}
\end{align}
Although the right-hand side depends on unknown coefficients $a^{(0)}_{\Om^-_{\ell,0}}$, one should not be too worried, since these can be calculated by the order $O(\veps^0)$ bootstrap, see section \ref{sec:PP2}.

\paragraph{Low-spin contributions}

There is an important caveat to the  dispersion relation \eqref{eq:disp-rel}, namely that it is correct provided $F(r,w)$ decays as $|w|^{-\delta}$ when $w \to \infty$, for some positive $\delta$.
If this decay does not hold, one needs to add extra terms to the result, that we call non-normalizable
\begin{align}
 F(r,w)
 = \DR \big[ F(r,w) \big]
 \, + \, \text{non-normalizable} \, .
 \label{eq:disp-rel-dots}
\end{align}
For example, whenever the defect identity appears in the defect expansion one gets a constant term $F_{\Om_1\Om_2} = a_{\Om_1} a_{\Om_2} + \ldots \, $.
Since this constant does not decay at infinity, it must be restored by hand.

In the rest of this work, we make the key assumption that the the dispersion relation only misses a constant contribution, or in other words, we are assuming the correlator goes as a constant as $|w| \to \infty$.
In more physical terms, we are assuming that the defect families have analytic CFT data all the way down to spin $s=0$.
This assumption is justified a posteriori for the localized magnetic field line defect, because the bootstrap prediction agrees with the diagrammatic calculation of section \ref{sec:feyndiag}.
However, it would be good to have a priori arguments for when this assumption is true or not, as we discuss in the conclusions.

In any case, the above assumption implies that the dispersion relation reconstructs the correlators up to constant pieces, which are identified with the one-point functions of $\phi$ and $\phi^2$:
\begin{align}
 F_{\phi\phi}(r,w)
 = \DR \big[ F_{\phi\phi}(r,w) \big]
 + a^2_{\phi} \, , \qquad
 F_{\phi\phi^2}(r,w)
 = \DR \big[ F_{\phi\phi^2}(r,w) \big]
 + a_{\phi} a_{\phi^2} \, .
 \label{eq:correct-dr}
\end{align}
Fortunately, these are also the free parameters that our correlators depended on from the start.
In total, the bootstrap reconstructs $F_{\phi\phi}$ and $F_{\phi\phi^2}$ up to the four real coefficients \eqref{eq:free-params}, but these will be completely fixed in section \ref{sec:free-coeffs}.

\subsection{Two-point function \texorpdfstring{$\vvev{ \phi \phi }$}{<< phi phi >>}}
\label{sec:disc-wf}

\subsubsection{Inversion of bulk identity}

Let us start applying the dispersion relation \eqref{eq:disp-rel} to the first term in \eqref{eq:sing-part-wf}, which corresponds to the bulk identity in the $\phi \times \phi$ OPE.
A simple calculation \cite{Barrat:2022psm,Bianchi:2022ppi} shows that this term gets mapped to itself:
\begin{align}
 \xi^{-\Dp}
 = \DR \left[ \xi^{-\Dp} \right] \, .
 \label{eq:dr-identity}
\end{align}

\subsubsection{Inversion of \texorpdfstring{$\phi^2$}{phi^2}}
\label{sec:inv-P2}

Next we consider the contribution of the energy operator in the bulk OPE $\phi \times \phi \sim \phi^2 + \ldots \,$, so we apply the dispersion relation to the second term in \eqref{eq:sing-part-wf}
\begin{align}
 H(r,w)
 \equiv
 \DR \left[
   \xi^{-1} \partial_{\Delta} f_{2, 0}(r,w) \right] \, .
 \label{eq:dr-P2}
\end{align}
First we compute the discontinuity of $\partial_{\Delta} f_{2, 0}(r,w)$ using the series expansion \eqref{eq:bulk-block-ser}.
We have not been able to express the result in closed form, but we can generate an expansion to high order in $z$ in the limit $z \to 0$. (Remember that $z = rw$ and $\zb = r/w$.)
Because the calculations are somewhat involved, we postpone technical details to appendix \ref{sec:disc}, and here just show the first few terms:
\begin{align}
\label{eq:disc-full}
   \frac{1}{2\pi i} \Disc
 & \left[ \xi^{-1} \partial_{\Delta} f_{2, 0}(r,w) \right]
   =  \notag \\
 & \qquad \frac{\sqrt{z} \left(\sqrt{\zb}-1\right)^2}{4 \zb} \left( 
      z (\zb+1)
    + z^2 \frac{\left(7 \zb+2 \sqrt{\zb}+7\right) \left(3 \zb^2+2 \zb+3\right)}{32 \zb} 
    + \ldots
 \right) \notag \\
 & \qquad - \frac{\sqrt{z}}{2} \log \frac{16 \zb }{z(1 + \sqrt{\zb})^4}
    \left( 
        1 
        + z \frac{(\zb+1)^2}{4 \zb}
        + z^2 \frac{\left(3 \zb^2+2 \zb+3\right)^2}{64 \zb^{2}}
        + \ldots
    \right) \, .
\end{align}
Instead of using the discontinuity in the dispersion relation \eqref{eq:disp-rel}, it is simpler to use the Lorentzian inversion formula \eqref{eq:inversion}.
Th output of the inversion formula is the CFT data of defect operators, which after resummation gives the same correlator as the integral transform $\DR$.

The inversion formula generates the defect CFT data twist-by-twist, where the defect twist is $\wh\Delta-s$.
For the lowest twist, we keep the leading term in $z$ in \eqref{eq:disc-full} and expand the integration kernel \eqref{eq:inversion} to leading order
\begin{align}
 b(\Dh,s)|_{\text{leading}}
 & = \frac{1}{4}
 \int_0^1 dz \int_1^\infty d\zb \,
 z^{-\frac{\Dh-s+1}{2}} \zb^{-\frac{\Dh+s+2}{2}}
 \left( \log z - \log \frac{16 \zb}{(1 + \sqrt{\zb})^4} \right) \, .
 \label{eq:leading-twist-int}
\end{align}
The $z$ and $\zb$ integrations can be computed in terms of analytically continued harmonic numbers $H_a$, and the result reads
\begin{align}
 b(\Dh,s)|_{\text{leading}}
 &  =
 -\frac{1}{s+1/2} \frac{1}{(\Dh-s-1)^2} \notag  \\
 & \quad
 - \frac{2}{\Dh-s-1}
 \left(\frac{H_{\frac{1}{2} (\Dh+s-1)}-H_{\frac{1}{2} (\Dh+s-2)}}{\Dh+s}
 -\frac{1}{(\Dh+s)^2}
 -\frac{1}{(2 s+1)^2}\right) 
 \, .
 \label{eq:leading-twist-fam}
\end{align}
The single and double poles at $\wh\Delta = s+1$ indicates this family has twist one, as expected.
The CFT data for higher-twist families $\Dh = s + 2n + 1$ follows from expanding the discontinuity and the inversion kernel to higher order in $z$.
Since the discontinuity is multiplied by $\veps$ we can evaluate the integration kernel exactly in $d=4$, which simplifies expressions.
At each order in $z$, the integrals are similar to \eqref{eq:leading-twist-int} and can be evaluated in closed form.
Even though single poles in $\Dh - (s + 2n + 1)$ are generated in this calculation, thanks to non-trivial cancellations their residues vanish.
Therefore, there is a single family of defect operators given by \eqref{eq:leading-twist-fam}.\footnote{Curiously, for monodromy defects in the Wilson-Fisher fixed point there is also a single defect family \cite{Gimenez-Grau:2021wiv}.}

The function \eqref{eq:leading-twist-fam} contains the defect CFT data, namely the residue corrects OPE coefficients, while the the double-pole gives anomalous dimensions, see \eqref{eq:poles-iform}.
So the correlator we are after is the sum
\begin{align}
\label{eq:expd-H}
 H(r,w)
 = \sum_{s=0}^\infty \left( 
    \frac{ H_s-H_{s-1/2} }{s+1/2}
  - \frac{ 1 }{(s+1/2)^2}
  + \frac{1}{s + 1/2} \, \partial_{\Dh}
 \right) \wh f_{s+1,s}(r,w) \, ,
\end{align}
where $\wh f_{\Dh,s}$ are the defect conformal blocks \eqref{eq:def-block}.
Using the formula for the blocks we obtain an elegant triple sum representation of $H(r,w)$:
\begin{align}
 H(r,w)
 = \sum_{n,s=0}^\infty \sum_{m=0}^s r^{2n+s+1} w^{2m-s} &
 \frac{\left(\frac{1}{2}\right)_m \left(\frac{1}{2}\right)_n (-s)_m (s+1)_n}{m! n! \left(\frac{1}{2}-s\right)_m \left(\frac{1}{2}+s\right)_{n+1}} \notag \\
 & \times \left(
      \log 4 r
    + H_{n+s}
    - H_{n+s+\frac{1}{2}}
    + H_{-\frac{1}{2}}
 \right) \, . \label{eq:FWF_defsum}
\end{align}
This expansion converges near $r = 0$, but we are also interested in the expansion around $r=1$, which determines the bulk block decomposition of $H(r,w)$.
By carefully rearranging the sums in \eqref{eq:FWF_defsum}, we could generate many terms of the $r=1$ expansion, which can then be compared to the series expansion of bulk blocks \eqref{eq:bulk-block-ser}.
Surprisingly, we find that the bulk-channel expansion of $H(r,w)$ contains a single conformal block:
\begin{align}
\label{eq:exp-H-PP}
 \xi \, H(r,w)
 & = \big(
    \partial_\Delta 
    - 1 - \log 2
 \big) f_{2,0}(r,w) \, .
\end{align}
Remember that this formula uses $\veps = 0$, which is allowed because $H(r,w)$ appears at $O(\veps)$ in the final correlator.
Even though it was somewhat complicated to derive \eqref{eq:exp-H-PP},\footnote{We can provide more details upon request, although the derivation of \eqref{eq:exp-H-PP} is not particularly elegant.} it is simple to check this identity numerically to high precision comparing \eqref{eq:FWF_defsum} and the series expansion of the bulk block \eqref{eq:scalar-block-exp}.
The two formulas also agree with the integral representation \eqref{eq:Hint} to be derived below.
Although we do not have a formula for $H(r,w)$ in terms of simple special functions, we have a good analytic handle of it, because we can use \eqref{eq:FWF_defsum} and \eqref{eq:exp-H-PP} to expand around $r=0$ and $r=1$ respectively, and the integral representation \eqref{eq:Hint} to evaluate numerically to high precision anywhere in $0 < r < 1$.
In practice, this is all we need to know about $H(r,w)$.

\subsection{Two-point function \texorpdfstring{$\vvev{ \phi \phi^2 }$}{<< phi phi2 >>}}
\label{sec:PP2}

\subsubsection{Inversion of \texorpdfstring{$\phi$}{phi}}
\label{sec:PP2-P}

The first contribution to $\vvev{\phi\phi^2}$ comes from the exchange of $\phi$ in the bulk OPE $\phi \times \phi^2$, which is somewhat subtle because of multiplet recombination.
In free theory the scalar $\phi$ furnishes a short representation of the conformal group, because the descendants generated from $\partial^2 \phi$ vanish by the equations of motion $\partial^2 \phi = 0$.
When we turn on interactions the equations of motion are $\partial^2 \phi \propto \veps \phi^3$, so $\phi^3$ becomes a descendant of $\phi$ \cite{Rychkov:2015naa}.
In terms of conformal blocks, multiplet recombination implies that $f_{\Dp,0}^{\Dp-\Dpp}$ admits an expansion with $\Delta=1$ and $\Delta=3$ blocks evaluated at free theory.
This expansion is derived in the appendix, see \eqref{eq:exp-Pblock}, and the part that contributes to the discontinuity reads
\begin{align}
 f_{\Dp,0}^{\Dp-\Dpp}(r,w)\Big|_{\Disc}
&= \xi^{\frac12\Dp}
 - \frac{\veps}{6} \, \xi^{1/2} \log \frac{4 r}{(1+r)^2}
 + \frac{3\veps}{4} \partial_\Delta f_{3,0}^{-1}
 + O(\veps^2) \, .
 \label{eq:blockP-disc}
\end{align}
The first term is a particular case of the bulk identity inversion \eqref{eq:dr-identity}.
The second term has extra $r$ dependence, but since the dispersion relation only depends on $w$, it does not have any effect.
All in all we get
\begin{align}
 \DR \left[ \xi^{-\frac12\Dpp}
 \right]
 = \xi^{-\frac12\Dpp} \, , \qquad
 \DR \left[ \xi^{-1} \log \frac{4 r}{(1+r)^2} \right]
 = \xi^{-1} \log \frac{4 r}{(1+r)^2} \, ,
\end{align}
where we multiplied the block by an overall factor $\xi^{-\frac12(\Dp+\Dpp)}$ as dictated by \eqref{eq:blk-block-exp}.
The third term in \eqref{eq:blockP-disc} can also be analysed with the dispersion relation:
\begin{align}
 G(r,w) \equiv \DR \left[\xi^{-3/2} \, \partial_\Delta f_{3,0}^{-1} \right] \, .
 \label{eq:Gdef}
\end{align}
It is not hard to obtain the discontinuity in closed form, which is essentially a logarithm. Integrating the discontinuity against the dispersive kernel \eqref{eq:disp-rel} one gets a formula with dilogarithms.
However, we argue below that $G(r,w)$ drops from the final result, so we shall not need this expression.

\subsubsection{Inversion of triple-twist operators}

Finally, we consider the contribution of triple-twist operators $\Om^-_{\ell,0} \sim \partial^\ell \phi^3$.
In this case the calculation of the discontinuity presents a new complication: we should first obtain the tree-level coefficients $\lambda^{(0)}_{\phi\phi^2\Om^-_{\ell,0}} a^{(0)}_{\Om^-_{\ell,0}}$,  multiply them with the anomalous dimension $\gamma_{\Om^-_{\ell,0}}= \frac{2}{3} \frac{(-1)^\ell}{\ell+1}$, and perform the infinite sum in \eqref{eq:sing-part-phi-phi2}.

We start obtaining the tree-level correlation function, which follows from inverting $\phi$ as in the previous section.
Setting $\veps = 0$ in \eqref{eq:blockP-disc} we see that only $\xi^{-1}$ needs to be inverted, but the result has to be complemented with the constant piece of \eqref{eq:correct-dr}.
The tree-level correlator then reads:
\begin{align}
 F_{\phi\phi^2}^{(0)}(r,w)
 = \lambda_{\phi\phi^2\phi}^{(0)} a_\phi^{(0)} \, \xi^{-1}
 + a_\phi^{(0)} a_{\phi^2}^{(0)} \, .
\end{align}
With the correlator at hand, we can perform a conformal block expansion as described at length in appendix \ref{sec:block-exp}:
\begin{align}
 \xi^{\frac32} F_{\phi\phi^2}^{(0)}(r,w)
 = \lambda_{\phi\phi^2\phi}^{(0)} a_\phi^{(0)} f^{-1}_{1,0}
   + a_\phi^{(0)} a_{\phi^2}^{(0)}
     \sum_{\substack{\ell=0\\ \ell \text{ even}}}^\infty 4^{-\ell} f_{3+\ell,\ell}^{-1} \, .
 \label{eq:FPP2-tree}
\end{align}
The first term in \eqref{eq:FPP2-tree} corresponds to the exchange $\phi\times\phi^2 \sim \phi$ and it has been considered in section \ref{sec:PP2-P}.
Here we are concerned with the infinite sum in \eqref{eq:FPP2-tree}, which contains the products of OPE coefficients that we need.
Multiplying them by the anomalous dimension we obtain the discontinuity to order $O(\veps)$.
Remember that the $\ell = 0$ term corresponds to $\Om^-_{\ell,0} \sim \phi^3$, and by multiplet recombination it should be absent, so we find
\begin{align}
 \Disc F_{\phi\phi^2}(r,w)|_{\Om^-_{\ell,0}}
 = \frac{2}{3} a_\phi^{(0)} a_{\phi^2}^{(0)} \, \veps
   \sum_{\substack{\ell=2\\ \ell \text{ even}}}^\infty \frac{4^{-\ell}}{\ell+1}
   \Disc \left[ \xi^{-3/2} \, \partial_\Delta f^{-1}_{3+\ell,\ell} \right] \, .
\end{align}
This infinite sum looks daunting at first, but we found that it is closely related to the function $H(r,w)$ studied in section \ref{sec:inv-P2}.
To be precise, if we include the $\ell = 0$ term in the sum we find
\begin{align}
 \Disc H(r,w)
 = \sum_{\substack{\ell=0\\ \ell \text{ even}}}^\infty \frac{4^{-\ell}}{\ell+1}
   \Disc \left[ \xi^{-3/2} \, \partial_\Delta f^{-1}_{3+\ell,\ell} \right] \, .
 \label{eq:relHinfsum}
\end{align}
Although we do not have an analytic proof of this formula, we observed empirically that the Taylor series of the two sides agree to the high order.
Alternatively, one can check this formula numerically to high precision.
All in all, we conclude that the triple-twist family contributes a term
\begin{align}
 F_{\phi\phi^2}(r,w)|_{\Om^-_{\ell,0}}
 = \frac{2}{3} a_\phi^{(0)} a_{\phi^2}^{(0)}\, \veps
   \big( H(r,w) - G(r,w) \big) \, ,
 \label{eq:dr-z2odd}
\end{align}
where we subtracted $G(r,w)$ defined by \eqref{eq:Gdef} to account for the extra $\ell=0$ term in \eqref{eq:relHinfsum}.

\subsection{Fixing the free coefficients}
\label{sec:free-coeffs}

In sections \ref{sec:disc-wf} and \ref{sec:PP2} we computed the building blocks for the correlation functions $F_{\phi\phi}$ and $F_{\phi\phi^2}$ respectively, while in section \ref{sec:struct-calc} we described how to assemble the different pieces.
For example, the correlator $F_{\phi\phi}$ is the combination of the identity and $\phi^2$ pieces given in \eqref{eq:dr-identity} and \eqref{eq:dr-P2} respectively, with the relative coefficients in \eqref{eq:sing-part-wf}:
\begin{align}
\label{eq:corr-final}
\begin{split}
  F_{\phi\phi}(r,w)
& = \xi^{-\Dp}
  + a_\phi^2
  + \frac{\veps}{3} \lambda_{\phi\phi\phi^2} a_{\phi^2}  H(r,w)
  + O(\veps^2) \, .
\end{split}
\end{align}
Similarly, the correlator $F_{\phi\phi^2}$ receives the contributions from $\phi$ in \eqref{eq:blockP-disc}-\eqref{eq:Gdef}, and a contribution from the triple-twist operators \eqref{eq:dr-z2odd}:
\begin{align}
\label{eq:corr-final-pp2}
  F_{\phi\phi^2}(r,w)
& = a_{\phi} a_{\phi^2}
  + \lambda_{\phi\phi\phi^2} a_\phi \left(
    \xi^{-\frac12 \Dpp}
  - \frac{\veps}{6 \xi} \log \frac{4r}{(1+r)^2}
  + \frac{3\veps}{4} \, G(r,w)
  \right) \notag \\[0.2em]
& \quad
  + \frac{2\veps}{3} a_\phi a_{\phi^2} \big( H(r,w) - G(r,w) \big)
  + O(\veps^2) \, .
\end{align}
These two correlators depend on CFT data that is known from bulk physics
\begin{align}
 \Dp
 = 1 - \frac{\veps}{2} + O(\veps^2) \, , \quad
 \Dpp
 = 2\Delta_\phi + \frac{\veps}{3} + O(\veps^2) \, , \quad
 \lambda_{\phi\phi\phi^2}
 = 2 - \frac{2\veps}{3} + O(\veps^2) \, ,
\end{align}
as well as the one-point function coefficients $a_\phi$ and $a_{\phi^2}$.
At order $O(\veps)$ the one-point coefficients correspond to four unknowns, see \eqref{eq:free-params}, which are fixed uniquely by the three physical requirements that we now discuss.

\paragraph{Existence of displacement.} The first consistency condition is that the defect must support a displacement operator $D^i$, that appears in the broken conservation equation for the stress tensor \cite{Billo:2016cpy}:
\begin{align}
\partial_\mu T^{\mu i} = C_D^{1/2} \, D^i(\hat x) \, \delta^{(q)}(\vec x) \, .
\end{align}
Note that we normalize $D^i$ to have unit two-point function, so we introduce an unknown constant $C_D$ in the Ward identity.
Importantly, the definition implies that the displacement operator has scaling dimension $\wh\Delta_D = 2$ and transverse-spin $s=1$.
We can then expand $F_{\phi\phi}$ in the defect channel using the results in appendix \ref{sec:app-def-exp}.
The contribution of the displacement reads:
\begin{align}
  F_{\phi\phi}(r,w)\Big|_{\substack{\wh\Delta=2 \\ s=1}}
& = \frac12 \left(2 - \veps
  + \frac{4 \sqrt{2}}{27} a_{\phi^2}^{(0)} (6 \log 2 - 5) \veps
  + \left(\frac{4 \sqrt{2}}{9} a_{\phi^2}^{(0)} - 1\right)
    \veps \, \partial_{\Dh}
  \right) \wh f_{2,1}(r,w) \, .
\end{align}
As we just mentioned, the displacement operator has fixed dimension $\wh\Delta_D = 2$, so the derivative term above must vanish.
This fixes the value of $a_{\phi^2}^{(0)}$, as well as the OPE coefficient $b_{\phi D}$ that we write for future reference
\begin{align}
 a_{\phi^2}^{(0)} = \frac{9}{4 \sqrt{2}} \, , \qquad
 b_{\phi D}^2
 = 2
 + \veps \log 4
 - \frac{8\veps}{3}
 + O(\veps^2)
  \, .
 \label{eq:ope-coeff-bpD}
\end{align}
At this point one notices that, as promised, the function $G(r,w)$ drops from $F_{\phi\phi^2}$ in equation \eqref{eq:corr-final-pp2}.

\paragraph{Displacement Ward identity.}

Besides having fixed dimension, the displacement operator satisfies a Ward identity that relates the coefficients $\Delta_\Om a_\Om \propto b_{\Om D}$, see \cite{Billo:2016cpy} for the derivation.
Since we do not have an independent way to determine $C_D$, we formulate the Ward identity as
\begin{align}
 \frac{\Delta_\phi a_\phi}{b_{\phi D}}
 = \frac{\Delta_{\phi^2} a_{\phi^2}}{b_{\phi^2 D}} \, ,
 \label{eq:wi}
\end{align}
which is manifestly invariant under rescaling of $D^i$.
To impose the Ward identity, we also need to expand $F_{\phi\phi^2}$ in the defect channel.
Using the value of $a_{\phi^2}^{(0)}$ just determined, the expansion gives a displacement with the correct scaling dimension
\begin{align}
  F_{\phi\phi^2}(r,w)\Big|_{\substack{\wh\Delta=2 \\ s=1}}
& = \frac12 \, b_{\phi D} b_{\phi^2 D} \wh f_{2,1}(r,w) \, ,
\end{align}
where the OPE coefficient reads
\begin{align}
  b_{\phi D} b_{\phi^2 D}
  = 2 \sqrt{2} a_\phi^{(0)} \left(1 - \frac{4 \veps}{3}
  + \frac{2}{3} \veps  \log 2 \right)
  + 2 \sqrt{2} a_\phi^{(1)} \veps
  + O(\veps^2) \, .
 \label{eq:ope-coeff-bp2D}
\end{align}
Plugging in the OPE coefficients \eqref{eq:ope-coeff-bpD} and \eqref{eq:ope-coeff-bp2D} in the Ward identity \eqref{eq:wi} fixes
\begin{align}
 a_\phi
 = \pm \left(
  \frac{3}{2}
  + \frac{\veps }{8}
  + \frac{\veps}{4} \log 2
  + \frac{\sqrt{2}}{3} a_{\phi^2}^{(1)} \veps
 \right) \, .
 \label{eq:rel-aP-aP2}
\end{align}
We have the freedom to choose either the positive or negative sign, since both are related by a redefinition of the field $\phi \to -\phi$.

\paragraph{Decoupling of $\phi^3$.}
The last condition on the CFT data follows from the bulk expansion.
In the case of $F_{\phi\phi}$, we require that $\phi^2$ appears with the correct OPE coefficient
\begin{align}
 \xi^{\Dp} F_{\phi\phi}(r,w)
 = 1
 + \lambda_{\phi\phi\phi^2} a_{\phi^2} f_{\Dpp,0}(r,w)
 + \ldots
\end{align}
However, this is automatically satisfied with the constraints imposed so far.
On the other hand, for $F_{\phi\phi^2}$ we should require that $\phi$ appears with the correct OPE coefficient, and that $\phi^3$ decouples by multiplet recombination:
\begin{align}
 \xi^{\frac12(\Dp+\Dpp)} F_{\phi\phi^2}(r,w)
&= \lambda_{\phi\phi\phi^2} a_{\phi} f_{\Dp,0}^{\Dp-\Dpp}
 + 0 \times f_{\Dp+\Dpp,0}^{\Dp-\Dpp}
 + \ldots \, .
 \label{eq:consPP2}
\end{align}
The results of appendix \ref{sec:block-exp} give the following expansion
\begin{align}
 \xi^{\frac12(\Dp+\Dpp)} F_{\phi\phi^2}(r,w)
&= - \frac{1}{24} \left(
     36 \sqrt{2}
   - 3 \sqrt{2} \veps
   + 6 \sqrt{2} \veps \log 2
   + 16 a_{\phi^2}^{(1)} \veps
  \right) f_{\Dp,0}^{\Dp-\Dpp} \notag \\
& \quad
 - \frac{\veps}{32} \left(
     17 \sqrt{2}
   - 36 \sqrt{2} \log 2
   + 48 a_{\phi^2}^{(1)}
 \right) f_{\Dp+\Dpp,0}^{\Dp-\Dpp}
 + \ldots \, ,
\end{align}
which should equal \eqref{eq:consPP2}.
This gives two equations, but one of them is redundant, and the remaining free parameter is fixed to
\begin{align}
 a_{\phi^2}^{(1)}
 = - \frac{17}{24 \sqrt{2}}
 + \frac{3 \log 2}{2 \sqrt{2}} \, .
\end{align}
As promised, the mixed correlator bootstrap combined with the above physical requirements determine uniquely the two-point functions.

\subsection{\texorpdfstring{$O(N)$}{O(N)} model}
\label{sec:on-model}

The bootstrap calculation generalizes naturally to line defects in the $O(N)$ model that break the symmetry down to $O(N-1)$.
Consider the two-point function of the order parameter:
\begin{align}
 \vvev{ \phi_a(x_1) \phi_b(x_2) }
&= \frac{F_{ab}(r,w)}{|\vec x_1|^{\Dp} |\vec x_2|^{\Dp}}  \, .
\label{eq:two-pt-ON}
\end{align}
As for the Ising model, we use information from the bulk to determine the functional form of $F_{ab}(r,w)$ up to a few parameters, that will be fixed from physical requirements.

The main difference compared to Ising is that $F_{ab}(r,w)$ decomposes into two global-symmetry channels.
The reason is the broken $O(N)$ symmetry, which allows for two different tensor structures $\delta_{ab}$ and $\delta_{a1} \delta_{b1}$.\footnote{We choose the $O(N-1)$ subgroup of rotations that preserve the vector $(1,0,0, \ldots)$.}
The OPE $\phi_a \times \phi_b$ consists of three global symmetry channels, the singlet $S$, the symmetric-traceless $T$, and the antisymmetric $A$.
However, the pattern of symmetry breaking $O(N)\to O(N-1)$ forces one-point functions of antisymmetric operators to vanish.
As a result, the OPE takes the following schematic form, where spin indices, descendants and dimensionful factors are neglected:
\begin{align}
 \lim_{x_{12}^2 \to 0} \phi_a(x_1) \phi_b(x_2)
 = \delta_{ab} \sum_{\Om \in S} \lambda_{\phi\phi\Om} \Om(x_2)
 + \mathbf T_{ab,cd} \sum_{\Om_{cd} \in T} \lambda_{\phi\phi\Om}
   \Om_{cd}(x_2) \, .
\end{align}
The object of interest here are the tensor structures, and we take the symmetric-traceless projector to be defined by
\begin{align}
 \mathbf{T}_{ab,cd}
 = \frac12 \delta_{ac} \delta_{bd}
 + \frac12 \delta_{ad} \delta_{bc}
 - \frac{1}{N} \delta_{ab} \delta_{cd} \, .
 \label{eq:def-Tabcd}
\end{align}
Inserting the OPE in the two-point function and defining $\vvev{\Om_{ab}} \sim a_\Om (\delta_{a1}\delta_{b1} - \frac1N \delta_{ab})$ for symmetric-traceless one-point functions, we find
\begin{align}
 F_{ab}(r,w)
&= \delta_{ab} F_{S}(r,w)
 + \left( \delta_{a1} \delta_{b1} - \frac{\delta_{ab}}{N} \right)
    F_{T}(r,w) \, .
  \label{eq:blk-tens-struct}
\end{align}
With this decomposition, $F_S$ and $F_T$ admit standard bulk-block expansions with the normalization conventions of appendix \ref{sec:review-dcft}.
Now we are ready to determine the form of the correlator dictated by bulk physics.
In the singlet channel the identity and $\phi^2$ operator can contribute to the discontinuity at order $O(\veps)$, while in the symmetric-traceless channel only the operator $T_{ab} = \phi_a \phi_b - \frac{\delta_{ab}}{N} \phi_c^2$ contributes.
Using the OPE coefficients and anomalous dimensions to be discussed in \eqref{eq:diml-p2}-\eqref{eq:diml-Tab}, the correlators must read
\begin{align}
 F_{S}(r,w)
&= \xi^{-\Dp}
 + \sqrt{\frac{2}{N}} \frac{N+2}{N+8} a_{\phi^2}^{(0)} \veps H(r,w)
 + A_S \, , \label{eq:ansantz-PP-ON} \\[0.2em]
 F_{T}(r,w)
&= \frac{2 \sqrt{2}}{N+8} a_{T_{ab}}^{(0)} \veps H(r,w)
 + A_T \, , \label{eq:ansantz-PP-ON-T}
\end{align}
where the constants $A_{S/T}$ are added in accordance with equation \eqref{eq:correct-dr}.

We would like to constrain the free parameters in \eqref{eq:ansantz-PP-ON} and \eqref{eq:ansantz-PP-ON-T} using defect information.
Note that the defect expansion consists on two channels $\wh S$ and $\wh V$, which stand for singlets and vectors under $O(N-1)$:
\begin{align}
 \lim_{|\vec x| \to 0} \phi_a(x)
 = \delta_{a1} \sum_{\wh\Om \in \wh S}
   b_{\phi\wh\Om} \, \wh \Om(\hat x)
 + \delta_{a\hat b} \sum_{\wh\Om \in \wh V} \wh \Om_{\hat b}(\hat x) \, .
\end{align}
Once again this should be inserted in the two-point function \eqref{eq:two-pt-ON}, giving the decomposition
\begin{align}
 F_{ab}(r,w)
&= \delta_{a1} \delta_{b1} F_{\wh S}(r,w)
 + (\delta_{ab} - \delta_{a1} \delta_{b1}) F_{\wh V}(r,w) \, .
 \label{eq:def-tens-struct}
\end{align}
The equivalence between \eqref{eq:blk-tens-struct} and \eqref{eq:def-tens-struct} is the crossing equation, which mixes the two bulk channels with the two defect channels.

We are finally in a position to fix the freedom in equations \eqref{eq:ansantz-PP-ON}-\eqref{eq:ansantz-PP-ON-T}.
On the one hand, the arguments of section \ref{sec:free-coeffs} imply the singlet channel contains the defect identity and the displacement operator.
The defect identity contributes a constant $a_\phi^2$, while the displacement contributes a block of dimension $\wh\Delta=2$ and transverse spin $s=1$.
A new feature of the $O(N)$ model is that there exists a defect operator, dubbed tilt operator in \cite{Padayasi:2021sik}, which captures the breaking of the global symmetry to $O(N-1)$.
In particular, the conservation of the current $\partial_\mu j_{[ab]}^\mu$ acquires a contact term on the defect
\begin{align}
 \partial_\mu j^\mu_{[1\hat a]}(x)
 = C_t^{1/2} t_{\hat a}(\hat x) \delta^{(q)}(\vec x) \, .
\end{align}
The tilt operator $t_{\hat a}$ has transverse spin $s=0$, dimension $\wh\Delta_t = 1$ and transforms as a vector under $O(N-1)$.
As a result, the low-lying terms in the defect expansion must have the form:
\begin{align}
 F_{\wh S}(r,w)
 = a_\phi^2
 + \frac12 b_{\phi D}^2 \wh f_{2,1}(r,w)
 + \ldots \, , \qquad
 F_{\wh V}(r,w)
 = b_{\phi t}^2 \wh f_{1,0}(r,w)
 + \ldots \, .
\end{align}
Demanding consistency of the correlators \eqref{eq:ansantz-PP-ON} and \eqref{eq:ansantz-PP-ON-T} with the defect expansion fixes
\begin{align}
 a^{(0)}_{\phi^2}
 = \frac{N+8}{4 \sqrt{2N}} \, , \qquad
 a_{T}^{(0)}
 = \frac{N+8}{4 \sqrt{2}} \, , \qquad
 A_S = \frac{1}{N} a_\phi^2 \, , \qquad
 A_T = a_\phi^2 \, .
 \label{eq:aP2-aT-boot}
\end{align}
The last condition to impose is consistency with the bulk channel decomposition
\begin{align}
 \xi^{\Dp} F_{S}(r,w)
 = \xi^{-\Dp}
 + \lambda_{\phi\phi\phi^2} a_{\phi^2} f_{\Delta_{\phi^2},0}
 + \ldots \, , \qquad
 \xi^{\Dp} F_{T}(r,w)
&= \lambda_{\phi\phi T} a_{T} f_{\Delta_{T},0}
 + \ldots \, ,
\end{align}
with the scaling dimensions and OPE coefficients given in \eqref{eq:diml-p2}-\eqref{eq:diml-Tab}.
Using appendix \ref{sec:block-exp} to carry out the bulk block expansion, we are able to fix
\begin{align}
 a_\phi^{(0)} = \pm \frac{\sqrt{N+8}}{2} \, ,
\end{align}
and to relate $a_{\phi}^{(1)}$ to the one-point functions of $\phi^2$ and $T_{ab}$:
\begin{align}
 a_{\phi^2}^{(1)}
&= \pm \sqrt{\frac{N+8}{2N}} \, a_\phi^{(1)}
 - \frac{N + 2 + (2 N +4) \log 2}{8 \sqrt{2N}} \, , \\
 a_{T}^{(1)}
&= \pm \sqrt{\frac{N+8}{2}} \, a_\phi^{(1)}
 - \frac{1+2 \log 2}{4 \sqrt{2}} \, .
\end{align}
However, since we also do not know $a_{\phi^2}^{(1)}$ and $a_{T}^{(1)}$, we are left with one free parameter in our results.
All in all, the full correlator reads
\begin{align}
  F_{ab}(r,w)
& = \delta_{ab} \, \xi^{-\Dp}
  + \delta_{a1} \delta_{b1} a_{\phi}^2
  + \frac{\veps}{4} \big( \delta_{ab} + 2\delta_{a1}\delta_{b1} \big) H(r,w)
  + O(\veps^2) \, ,
\label{eq:Fab-boot}
\end{align}
where $a_\phi^{(0)} = \pm \frac{\sqrt{N+8}}{2} + a_{\phi}^{(1)} \veps$.
Presumably, if we could bootstrap $F_{\phi\phi^2}$ to order $O(\veps)$ we would fix the last free parameter, as in section \ref{sec:free-coeffs}.
However, when $N>1$ the calculation becomes more difficult, because the triple-twist family $\Om_{\ell,0}^V \sim \partial^\ell \phi^2 \phi_a$ contains two operators per spin with anomalous dimension.\footnote{We thank Johan Henriksson for very useful clarification on this point.}
The bootstrap of $F_{\phi\phi^2}$ to order $O(\veps)$ requires to first solve this mixing problem, and we do not attempt to do this in the present work.

\section{Diagrammatic calculation}
\label{sec:feyndiag}

In this section, we use diagrammatic perturbation theory for the localized magnetic line defect in the $O(N)$ Wilson-Fisher model, see equation \eqref{eq:def-defect}.
We determine previously unknown CFT data, including bulk one-point functions and bulk-defect two-point functions.
We also determine bulk two-point functions, extending the bootstrap analysis of section \ref{sec:bootstrap}, and finding agreement when comparison is possible.
We follow closely the conventions and notation from \cite{Cuomo:2021kfm} (see also \cite{Gimenez-Grau:2022czc}), where similar calculations were performed.

\subsection{Conventions}

We consider the Wilson-Fisher $O(N)$ model in $d = 4-\veps$ dimensions.
The fundamental field is a scalar $\phi_{a}$ for $a = 1, \ldots, N$, described by the following action:\footnote{The notation $\phi_{0a}$ distinguishes the bare field in the Lagrangian from the renormalized one $\phi_a$, so the zero should not be thought of as an extra index.}
\begin{align}
 S
 = \int d^d x \left(
   \frac12 (\partial_\mu \phi_{0a})^2
 + \frac{\lambda_0}{4!} (\phi_{0a}^2)^2 \right)
 + h_0 \int_{-\infty}^\infty d\tau \, \phi_{01}(x(\tau)) \, .
 \label{eq:lmld-action}
\end{align}
This Lagrangian depends on the bare field $\phi_{0a}$ and bare couplings $\lambda_0$ and $h_0$.
The renormalized couplings $\lambda$ and $h$ are fixed by cancelling divergences from observable quantities.
With dimreg in the minimal-subtraction MS scheme, and working to leading order in $\lambda$, the renormalized couplings read \cite{Cuomo:2021kfm}
\begin{align}
 \lambda_0 = \lambda M^\veps \left(
  1 + \frac{\lambda}{(4\pi)^2} \frac{N+8}{3\veps} + O(\lambda^2)
 \right) \, , \quad
 h_0 = h M^{\veps/2} \left(
  1 + \frac{\lambda}{(4\pi)^2} \frac{h^2}{12\veps} + O(\lambda^2)
 \right) \, ,
 \label{eq:renorm-coupl}
\end{align}
where we introduce the arbitrary renormalization scale $M$.
As usual, since the bare couplings are independent of $M$, the renormalized couplings must depend on $M$, as shown by the beta functions.
In $d=4-\veps$ dimensions, there exists a non-trivial IR fixed point, which is reached at the critical couplings
\begin{align}
 \frac{\lambda_*}{(4\pi)^2}
 = \frac{3\veps}{N+8} + O(\veps) \, , \qquad
 h_*^2
 = N+8
 + \frac{4 N^2+45 N+170}{2 N+16} \, \veps
 + O(\veps^2) \, .
 \label{eq:critical-coupling}
\end{align}
In the calculations below, we present only results evaluated at this critical point.

Besides the couplings, we also need to renormalize the fundamental and composite fields.
For example, the renormalized energy operator is $\phi_{0}^2 = \phi_{0a} \phi_{0a} = Z_{\phi^2} \phi^2$,\footnote{Reference \cite{Cuomo:2021kfm} uses $\Om$ for bare operators and $[\Om]$ for renormalized ones. Instead, we call bare operators $\Om_0$ and renormalized operators $\Om$, but we hope this does not create confusion.}
and its two-point function is normalized as $\vev{\phi^2 \phi^2} \propto \Nm_{\phi^2}^2$.
For the operators $\phi^2$ and $T_{ab}$, both the renormalization factor and two-point function can be found in \cite{Cuomo:2021kfm}.
Similarly, we also need to renormalize defect fields and compute their two-point functions, which we take from \cite{Gimenez-Grau:2022czc}.

For perturbation theory, we need the propagator of the fundamental field when all couplings are zero:
\begin{align}
  \begin{tikzpicture}[baseline,valign]
  \draw[dashed] (0,  0) -- (1, 0);
\end{tikzpicture}
 \; \equiv \;
 \langle \phi_{0a}(x_1) \phi_{0b}(x_2) \rangle_{\lambda_0=h_0=0}
 = \frac{\kappa \delta_{ab}}{(x_{12}^2)^{1-\frac\veps 2}} \, , \qquad
 \kappa
 = \frac{\Gamma\!\left(\frac d2\right)}{2\pi^{d/2} (d-2)} \, .
 \label{eq:prop-lmld}
\end{align}
Furthermore, to build Feynman diagrams we have to use the vertices in the action \eqref{eq:lmld-action}
\begin{align}
  \begin{tikzpicture}[baseline,valign]
  \draw[dashed] (0,  0) -- (.5, .5);
  \draw[dashed] (0, .5) -- (.5,  0);
  \node at (0.25, .25) [bcirc] {};
\end{tikzpicture}
 \; \equiv \;
 - \lambda_0 \int d^d x \, \ldots \, , \qquad
 \begin{tikzpicture}[baseline,valign]
  \draw[thick] (0,  0) -- (.8, 0);
  \draw[dashed] (0.4, .6) -- (.4,  0);
  \node at (0.4, 0) [dcirc] {};
\end{tikzpicture}
 \; \equiv \;
 - h_0 \int_{-\infty}^\infty d\tau \, \ldots \, .
\end{align}
The first is the usual four-point vertex, while in the second the defect ``eats'' a bulk field.
Solid dots are always integrated, while lines that end without a dot correspond to points that are not integrated.

We are interested in computing correlators $\vvev{\Om_1 \ldots \wh \Om_1 \ldots }$, where the double-bracket notation reminds that we use the action  \eqref{eq:lmld-action}, which includes the line defect.
In a few occasions we also need correlators $\vev{\Om_1 \ldots }$, where the single-bracket indicates we set the defect action to zero $h_0 = 0$.
We shall compute these observables at most up to order $O(\veps)$.
Because the critical couplings are $h_* \sim O(1)$ and $\lambda_* \sim O(\veps)$, we must include all diagrams with at most one four-point vertex, but an arbitrary number of defect insertions.
Fortunately, for any given observable only a finite number of diagrams is required.

\subsection{Bulk one-point functions}

\subsubsection{Operator \texorpdfstring{$\phi_a$}{phi}}

Let us start with the simplest observable, the one-point function of the fundamental field.
Note that the renormalization factor is $Z_\phi = 1 + O(\lambda^2)$, so to the order we work, we do not need to distinguish the bare and renormalized fundamental fields $\phi_a = \phi_{0a}$.
In particular, its two-point function \eqref{eq:prop-lmld} is normalized as $\vev{\phi_a \phi_b} \sim \Nm_{\phi}^2$, where we introduce $\Nm_\phi^2 = \kappa$.
We then write the one-point function as
\begin{align}
 \vvev{ \phi_a(x) }
 = \frac{\delta_{a1} \, \Nm_{\phi} \, a_\phi}{|\vec x|^{\Dp}} \, ,
 \label{eq:form-aP}
\end{align}
so the coefficient $a_\phi$ is defined with the same normalization conventions as in \eqref{eq:one-pt-fun}.

Up to order $O(\veps)$, only two diagrams contribute:
\begin{align}
 \vvev{ \phi_a(x) }
 \quad = \quad
 \begin{tikzpicture}[baseline,valign]
  \draw[thick]  ( 0,  0) -- ( 1.2, 0);
  \draw[dashed] (.6, .9) -- (.6, 0);
  \node at (0.6, 0) [dcirc] {};
\end{tikzpicture}
\quad + \quad
 \begin{tikzpicture}[baseline,valign]
  \draw[thick]  ( 0,  0) -- ( 1.2, 0);
  \draw[dashed] (.6, .9) -- (.6, 0);
  \draw[dashed] (.3, 0) -- (.6, .5) -- (.9, 0);
  \node at (0.3, 0.0) [dcirc] {};
  \node at (0.6, 0.0) [dcirc] {};
  \node at (0.9, 0.0) [dcirc] {};
  \node at (0.6, 0.5) [bcirc] {};
\end{tikzpicture}
\quad + \quad \ldots
\end{align}
The computation of the first diagram involves the following elementary integral
\begin{align}
 \begin{tikzpicture}[baseline,valign]
  \draw[thick]  ( 0,  0) -- ( 1.2, 0);
  \draw[dashed] (.6, .9) -- (.6, 0);
  \node at (0.6, 0) [dcirc] {};
\end{tikzpicture}
 \quad \propto \quad
 \int_{-\infty}^\infty
     \frac{d\tau'}{\big[\vec x^2 + (\tau-\tau')^2 \big]^{1-\veps/2}}
  \; = \;
  \frac{\pi^{1/2} \, \Gamma \! \left(\frac{1-\veps}{2}\right)}
       {\Gamma \! \left(1-\frac{\veps}{2}\right)}
    \frac{1}{|\vec x|^{1-\veps}} \, .
 \label{eq:def-integral}
\end{align}
For the second diagram, we integrate four times in the $\tau$ directions using \eqref{eq:def-integral}.
We are left with a ``chain integral'' over $d-1$ dimensions, which can be evaluated as
\begin{align}
 \int \frac{d^{d-1} \vec y}{|\vec x - \vec y|^{a} |\vec y|^{b}}
 =
 \frac{\Gamma \! \left(\frac{d-1-a}{2}\right)
       \Gamma \! \left(\frac{d-1-b}{2}\right)
       \Gamma \! \left(\frac{a+b+1-d}{2}\right)}{
       \Gamma \! \left( \frac a2 \right)
       \Gamma \! \left( \frac b2 \right)
       \Gamma \! \left( \frac{2d-2-a-b}{2} \right)}
  \frac{\pi^{\frac{d-1}{2}}}{|\vec x|^{a+b+1-d}} \, .
\label{eq:chain-integral}
\end{align}
All that is left is to multiply the two diagrams with symmetry factors and relate the bare couplings to the renormalized ones as in \eqref{eq:renorm-coupl}. It is not hard to see that the divergencies cancel as expected.
We then evaluate the results at the critical point \eqref{eq:critical-coupling}, and compare with \eqref{eq:form-aP} to extract the one-point function \cite{Cuomo:2021kfm}
\begin{align}
 a_\phi
&= - \frac{\sqrt{N+8}}{2} \left[
   1
 + \veps \, \frac{(N+8)^2 \log 4 + N^2 - 3N - 22}{4 (N+8)^2}
 + O(\veps^2) \right] \, .
\label{eq:val-aP}
\end{align}
The sign of $a_\phi$ does not have a physical meaning, because it can be changed redefining $\phi \to -\phi$, or equivalently by redefining the critical coupling $h_* \to - h_*$.
Here and below we chose the $h_* > 0$ solution in \eqref{eq:critical-coupling}, which makes $a_\phi < 0$.

\subsubsection{Operators \texorpdfstring{$\phi^2$}{phi2} and \texorpdfstring{$T_{ab}$}{Tab}}

Using the same techniques, we can compute one-point functions of several other interesting operators.
For example, we can consider the $O(N)$-singlet and symmetric-traceless scalars
\begin{align}
 Z_{\phi^2} \phi^2 = \phi_{0a}^2 \, , \qquad
 Z_{T} T_{ab}
 = \phi_{0a} \phi_{0b} - \frac{\delta_{ab}}{N} \phi_{0c}^2 \, .
\end{align}
The renormalization factors $Z_\Om$, as well as the normalization of their two-point functions can be found for example in \cite{Cuomo:2021kfm}.
Due to the $O(N)$-symmetry, the one-point functions read
\begin{align}
 \vvev{ \phi^2(x) }
 = \frac{\Nm_{\phi^2} \, a_{\phi^2}}{|\vec x|^{\Delta_{\phi^2}}} \, , \qquad
 \vvev{ T_{ab}(x) }
 = \left( \delta_{a1} \delta_{b1} - \frac{\delta_{ab}}{N} \right)
 \frac{\Nm_{T} \, a_{T}}{|\vec x|^{2\Delta_T}} \, .
\end{align}
Once again, it is not hard to list all diagrams up to $O(\veps)$
\begin{align}
 \vvev{ \phi^2(x) }
 \quad = \quad
 \begin{tikzpicture}[baseline,valign]
  \draw[thick]  ( 0,  0) -- ( 1.2, 0);
  \draw[dashed] (.3,  0) -- (.6, .9) -- (.9, 0);
  \node at (0.3, 0) [dcirc] {};
  \node at (0.9, 0) [dcirc] {};
\end{tikzpicture}
\quad + \quad
 \begin{tikzpicture}[baseline,valign]
  \draw[thick]  (-.3,  0) -- ( 1.2, 0);
  \draw[dashed] ( .4, .9) -- ( .6, .5) -- (.6, 0);
  \draw[dashed] ( .4, .9) -- (0, 0);
  \draw[dashed] ( .3, 0) -- (.6, .5) -- (.9, 0);
  \node at (0.0, 0.0) [dcirc] {};
  \node at (0.3, 0.0) [dcirc] {};
  \node at (0.6, 0.0) [dcirc] {};
  \node at (0.9, 0.0) [dcirc] {};
  \node at (0.6, 0.5) [bcirc] {};
\end{tikzpicture}
\quad + \quad
\begin{tikzpicture}[baseline,valign]
  \draw[thick]  ( 0,  0) -- ( 1.2, 0);
  \draw[dashed] (.6, .9) to[out=-120,in=120] (.6, 0.3);
  \draw[dashed] (.6, .9) to[out=- 60,in= 60] (.6, 0.3);
  \draw[dashed] (.3, 0) -- (.6, .3) -- (.9, 0);
  \node at (0.3, 0.0) [dcirc] {};
  \node at (0.9, 0.0) [dcirc] {};
  \node at (0.6, 0.3) [bcirc] {};
\end{tikzpicture}
\quad + \quad \ldots
\label{eq:diags-phi2}
\end{align}
Exactly the same diagrams contribute for $T_{ab}$, except the corresponding symmetry factors will be different.
The integrals in the diagrams \eqref{eq:diags-phi2} can be calculated using \eqref{eq:def-integral} and \eqref{eq:chain-integral} repeatedly.
After accounting for the renormalization factors and evaluating at the critical point, one finds \cite{Cuomo:2021kfm}
\begin{align}
 a_{\phi^2}
&= \frac{N+8}{4\sqrt{2N}} \left[ 
   1
 + \veps \, \frac{12(N+8) \log 2 - 13 N - 38}{2(N+8)^2}
 + O(\veps^2)
 \right] \, , \label{eq:aP2-val} \\
 a_T
&= \frac{N+8}{4 \sqrt{2}} \left[ 
   1
 + \veps \, \frac{(N+6)(N+8) \log 4 + N^2 - 5N - 38}{2(N+8)^2}
 + O(\veps^2)
 \right] \, .
 \label{eq:aT-val}
\end{align}
At this point, we can look back and compare with the bootstrap results.
In particular, setting $N=1$ we see equation \eqref{eq:rel-aP-aP2} is satisfied.
For general $N$, we also find agreement between the tree-level part of \eqref{eq:aP2-val} and \eqref{eq:aT-val}, and the bootstrap result \eqref{eq:aP2-aT-boot}.

\subsubsection{Operators \texorpdfstring{$\phi^4$}{phi4} and \texorpdfstring{$\phi^2 T_{ab}$}{phi2 Tab}}

Finally, let us compute one-point functions of operators $\phi^4$ and $\phi^2 T_{ab}$.
We compute these operators at tree-level because our bootstrap analysis cannot probe higher order corrections.
However, let us emphasize that the computation of these corrections does not pose any new problems compared to the examples presented above.
First we need the normalization of the two-point functions
\begin{align}
 \langle \phi^4(x_1) \, \phi^4(x_2) \rangle
 & = \frac{\Nm_{\phi^4}^2}{(x_{12}^2)^{\Delta_{\phi^4}}}
 \, , \qquad \quad \;\;
 \Nm_{\phi^4}^2 = 8 \kappa^4 N (N+2) + O(\veps) \, , \\
 \langle \phi^2 T_{ab} (x_1) \, \phi^2 T_{cd}(x_2) \rangle
 & = \frac{\mathbf{T}_{ab,cd} \, \Nm_{\phi^2 T}^2}
          {(x_{12}^2)^{\Delta_{\phi^2 t}}}
 \, , \qquad
 \Nm_{\phi^2 T}^2 = 4 \kappa^4 (N+4) + O(\veps) \, .
\end{align}
The symmetric-traceless tensor structure $\mathbf{T}_{ab,cd}$ is defined in \eqref{eq:def-Tabcd}.
Since we work at tree level, only one diagram contributes to the one-point functions:
\begin{align}
 \vvev{ \phi^2 \phi_a \phi_b(x) }
 \quad = \quad
 \begin{tikzpicture}[baseline,valign]
  \draw[thick]  ( 0,  0) -- ( 1.2, 0);
  \draw[dashed] (.2,  0) -- (.6, .9) -- (1, 0);
  \draw[dashed] (.45,  0) -- (.6, .9) -- (.75, 0);
  \node at (0.20, 0) [dcirc] {};
  \node at (0.45, 0) [dcirc] {};
  \node at (0.75, 0) [dcirc] {};
  \node at (1.00, 0) [dcirc] {};
\end{tikzpicture}
\quad + \quad \ldots \, .
\end{align}
It can be readily evaluated with methods described above, and separating the two irreducible representations gives
\begin{align}
 a_{\phi^4}
 = \frac{(N+8)^2}{32 \sqrt{2} \sqrt{N (N+2)}} 
 + O(\veps) \, , \qquad
 a_{\phi^2 T}
 = \frac{(N+8)^2}{32 \sqrt{N+4}}
 + O(\veps) \, .
 \label{eq:aP4}
\end{align}
In section \ref{sec:ope-analysis} we show that these one-point functions are also in perfect agreement with the predictions from our bootstrap calculation.

\subsection{Bulk-defect two-point functions}

We now consider bulk-defect correlators, a type of observable that was not previously considered in \cite{Cuomo:2021kfm,Gimenez-Grau:2022czc}.
On the defect, we shall focus on the lowest-lying scalars, namely $\wh \phi_1$ and the tilt operator $t_{\hat a}$:
\begin{align}
 \phi_{01}(\tau)
 = Z_{\wh\phi_1} \, \wh \phi_1(\tau) \, , \qquad
 \phi_{0\hat a}(\tau)
 = Z_{t} \, t_{\hat a}(\tau) \, .
\end{align}
The notation $\Om(\tau)$ implies we are setting the perpendicular directions to zero $\vec x = 0$.
If we choose the bulk operator to be the fundamental scalar $\phi_a$, then there are two diagrams at order $O(\veps)$:
\begin{align}
 \vvev{ \phi_a(x) \wh \phi_b(\tau) }
  \quad = \quad
 \begin{tikzpicture}[baseline,valign]
  \draw[thick]  ( 0,  0) -- ( 1.2, 0);
  \draw[dashed] (.6, .9) -- (.6, 0);
\end{tikzpicture}
\quad + \quad
 \begin{tikzpicture}[baseline,valign]
  \draw[thick]  ( 0,  0) -- ( 1.2, 0);
  \draw[dashed] (.6, .9) -- (.6, 0);
  \draw[dashed] (.3, 0) -- (.6, .5) -- (.9, 0);
  \node at (0.3, 0.0) [dcirc] {};
  \node at (0.9, 0.0) [dcirc] {};
  \node at (0.6, 0.5) [bcirc] {};
\end{tikzpicture}
\quad + \quad \ldots
\end{align}
The first diagram does not involve integration so it does not pose further problems.
The second diagram contains a non-trivial integral
\begin{align}
 \begin{tikzpicture}[baseline,valign]
  \draw[thick]  ( 0,  0) -- ( 1.2, 0);
  \draw[dashed] (.6, .9) -- (.6, 0);
  \draw[dashed] (.3, 0) -- (.6, .5) -- (.9, 0);
  \node at (0.3, 0.0) [dcirc] {};
  \node at (0.9, 0.0) [dcirc] {};
  \node at (0.6, 0.5) [bcirc] {};
\end{tikzpicture}
  \quad & \propto \;
  \int \frac{M^{2\veps} \, d\tau_3 \, d\tau_4 \, d^d x_5}
            {\big[
             (\tau_{15}^2 + \vec x_{15}^2)
             (\tau_{25}^2 + \vec x_{5}^2)
             (\tau_{35}^2 + \vec x_{5}^2)
             (\tau_{45}^2 + \vec x_{5}^2) \big]^{(d-2)/2}}
             \notag \\[0.2em]
& = \frac{\pi ^4}{\big(\vec x_1^2 + \tau_{12}^2\big)^{1-\veps/2}} \left[
    \frac{2}{\veps }
  + \log \frac{M^4 \left(\vec x_1^2 + \tau_{12}^2\right)^4}
              {16 \, \vec x_1^4}
  + 7
  - \aleph
  + O(\veps)
  \right] \, ,
  \label{eq:bdd-hard-int}
\end{align}
where we introduce $\aleph = 1 + \gamma_E + \log \pi$.
To compute this diagram, we have first integrated over $\tau_3$ and $\tau_4$ using \eqref{eq:def-integral}.
We then introduce Schwinger parameters and integrate over $\tau_5$ and $\vec x_5$.
After using the analytic regularization of \cite{Panzer:2014gra,Panzer:2015ida}, which is also implemented in \texttt{HyperInt} \cite{Panzer:2014caa}, we expand to order $O(\veps^0)$.
The resulting integrals are somewhat complicated, but can be evaluated in \mathematica.

The form of these correlators is fixed by conformal symmetry
\begin{align}
 \vvev{ \phi_a(x_1) \, \wh \phi_1(\tau_2) }
 = \frac{\delta_{a1} \, \Nm_\phi \, \Nm_{\wh \phi_1} \,
         b_{\phi\,\wh\phi_1}}
        {(|\vec x_1|^2 + \tau_{12}^2)^{\wh\Delta_{\wh\phi_1}}
          |\vec x_1|^{\Dp-\wh\Delta_{\wh\phi_1}}} \, , \quad
 \vvev{ \phi_a(x_1) \, t_{\hat b}(\tau_2) }
 = \frac{\delta_{a\hat b} \, \Nm_\phi \, \Nm_{t} \, b_{\phi \, t}}
        {(|\vec x_1|^2 + \tau_{12}^2) |\vec x_1|^{\Dp-1}} \, ,
\end{align}
where the scaling dimensions are known $\wh\Delta_{\wh\phi_1} = 1 + \veps + O(\veps^2)$ and $\wh\Delta_t = 1$.
The renormalization factors and two-point normalizations are also known \cite{Gimenez-Grau:2022czc}
\begin{align}
\begin{split}
 Z_{\wh\phi_1}
 & = 1 - \frac{\lambda}{(4\pi)^2} \frac{h^2}{4\veps} + O(\lambda^2) \, , \\
 \Nm_{\wh\phi_1}^2
 & = \kappa \left(1 - \frac{3\aleph}{2} \, \veps + O(\veps^2) \right) \, ,
\end{split} \quad
\begin{split}
 Z_{t}
 & = 1 - \frac{\lambda}{(4\pi)^2} \frac{h^2}{12\veps} + O(\lambda^2) \, , \\
 \Nm_{t}^2
 & = \kappa \left(1 - \frac{\aleph}{2} \, \veps + O(\veps^2) \right) \, .
\end{split}
\end{align}
Combining this information and evaluating at the critical point, we find
\begin{align}
 b_{\phi \, \wh \phi_1}
 = 1 - \frac{3-3\log 2}{2} \, \veps + O(\veps^2) \, , \qquad
 b_{\phi \, t}
 = 1 - \frac{1- \log 2}{2} \, \veps + O(\veps^2) \, .
 \label{eq:b-diag}
\end{align}

\subsection{Bulk-bulk two-point functions}

The last observable we compute are two-point functions of the bulk scalars $\phi_a$ and $\phi^2$.
The results here extend the bootstrap calculation by determining $\vvev{\phi \phi^2}$, which was unknown for $N>1$, and by determining $\vvev{\phi^2 \phi^2}$.
Both of these results would have been dificult to obtain with bootstrap methods, due to mixing between nearly-degenerate operators.
Besides checking the bootstrap calculation, these results give access to CFT data for infinite families of operators, as we discuss in section \ref{sec:ope-analysis}.

\subsubsection{Correlator \texorpdfstring{$\vvev{\phi_a \phi_b}$}{<phi phi>}}

The calculation of two-point functions proceeds identically to the previous sections.
We start by listing all diagrams that contribute up to $O(\veps)$:
\begin{align}
 \vvev{ \phi_a(x_1) \phi_b(x_2) }
 \quad = \quad
\begin{tikzpicture}[baseline,valign]
  \draw[thick]  ( 0,  0) -- ( 1.2, 0);
  \draw[dashed] (0.2, 0.7) -- (1.0, 0.7);
\end{tikzpicture}
\quad + \quad
 \begin{tikzpicture}[baseline,valign]
  \draw[thick]  ( 0,  0) -- ( 1.2, 0);
  \draw[dashed] (.3, .9) -- (.3, 0);
  \draw[dashed] (.9, .9) -- (.9, 0);
  \node at (0.3, 0) [dcirc] {};
  \node at (0.9, 0) [dcirc] {};
\end{tikzpicture}
\quad + \quad
 \begin{tikzpicture}[baseline,valign]
  \draw[thick]  ( 0,  0) -- ( 1.2, 0);
  \draw[dashed] (.3, .9) -- (.9, 0);
  \draw[dashed] (.9, .9) -- (.3, 0);
  \node at (0.3, 0) [dcirc] {};
  \node at (0.9, 0) [dcirc] {};
  \node at (0.6, 0.45) [bcirc] {};
\end{tikzpicture}
\quad + \quad
 \begin{tikzpicture}[baseline,valign]
  \draw[thick]  ( 0,  0) -- ( 1.5, 0);
  \draw[dashed] (.6, .9) -- (.6, 0);
  \draw[dashed] (.3, 0) -- (.6, .5) -- (.9, 0);
  \draw[dashed] (1.2, .9) -- (1.2, 0);
  \node at (0.3, 0.0) [dcirc] {};
  \node at (0.6, 0.0) [dcirc] {};
  \node at (0.9, 0.0) [dcirc] {};
  \node at (1.2, 0.0) [dcirc] {};
  \node at (0.6, 0.5) [bcirc] {};
\end{tikzpicture}
\quad + \quad \ldots
\label{eq:diags-PP}
\end{align}
The first diagram is just the free-theory propagator, while the second and fourth diagrams renormalize the one-point function of $\phi_a$.
The only non-trivial diagram at this order is the third.
Since the diagram is finite, and it comes multiplied with a coupling $\lambda_* \sim O(\veps)$, it suffices to evaluate the diagram in $d=4$.
Interestingly, this diagram corresponds exactly to the function $H(r,w)$ introduced in \eqref{eq:dr-P2}:\footnote{This integral was also considered in \cite{Barrat:2020vch}, where it was called the ``$X$-integral''.
In that work, this integral was expanded numerically, while in our case, the bootstrap provided two natural expansions, see  \eqref{eq:FWF_defsum}-\eqref{eq:exp-H-PP}. We thank Pedro Liendo for pointing this out.}
\begin{align}
 \begin{tikzpicture}[baseline,valign]
  \draw[thick]  ( 0,  0) -- ( 1.2, 0);
  \draw[dashed] (.3, .9) -- (.9, 0);
  \draw[dashed] (.9, .9) -- (.3, 0);
  \node at (0.3, 0) [dcirc] {};
  \node at (0.9, 0) [dcirc] {};
  \node at (0.6, 0.45) [bcirc] {};
\end{tikzpicture}
 \quad \propto \quad
 \int \frac{d\tau_3 \, d\tau_4 \, d^4 x_5}
           {(\tau_{15}^2 + \vec x_{15}^2)
            (\tau_{25}^2 + \vec x_{25}^2)
            (\tau_{35}^2 + \vec x_{5}^2)
            (\tau_{45}^2 + \vec x_{5}^2)}
 \equiv - 2 \pi ^4 \frac{H(r,w)}{|\vec x_1| |\vec x_2|} \, .
\end{align}
The easiest way to evaluate the integral is to first integrate $\tau_3$ and $\tau_4$, then introduce Schwinger parameters to integrate over $x_5$, and finally integrate over the Schwinger parameters.
At the end, we are left with one last integral that we do not know how to evaluate:
\begin{align}
 H(r,w)
 = - \int_0^\infty d \alpha \,
 \sqrt{ \frac{z \zb}{(\alpha+1) (\alpha+z \zb) (\alpha+z) (\alpha+\zb)} } \,
 \tanh^{-1} \sqrt{\frac{(\alpha+z) (\alpha+\zb)}{(\alpha+1) (\alpha+z \zb)}} \, .
 \label{eq:Hint}
\end{align}
However, we can check numerically to high precision that this integral agrees with the series expansion \eqref{eq:FWF_defsum} and with \eqref{eq:exp-H-PP}.
In practice, this means we know both the bulk- and defect-channel expansions of $H(r,w)$.

In order to obtain the correlator in the standard CFT normalization, we combine the diagrams in \eqref{eq:diags-PP}, divide by $\Nm_\phi^2$, and multiply by the overall factor $(|\vec x_1| |\vec x_2|)^{\Dp}$.
All in all we find
\begin{align}
  F_{ab}(r,w)
& = \delta_{ab} \, \xi^{-\Dp}
  + \delta_{a1} \delta_{b1} a_{\phi}^2
  + \frac{\veps}{4} \big( \delta_{ab} + 2\delta_{a1}\delta_{b1} \big) H(r,w)
  + O(\veps^2) \, ,
 \label{eq:PP-diag}
\end{align}
where $a_\phi$ takes the value reported in \eqref{eq:val-aP}.
Once again, we find perfect agreement with the bootstrap result \eqref{eq:Fab-boot}.

\subsubsection{Correlator \texorpdfstring{$\vvev{\phi_a \phi^2}$}{<phi phi2>}}

In a similar way, one can obtain the two-point function of the fundamental field and the energy operator.
In this case the diagrams that contribute are
\begin{align}
 \vvev{ \phi_a(x) \phi^2(x) }
 \quad & = \quad
 \begin{tikzpicture}[baseline,valign]
  \draw[thick]  ( 0,  0) -- ( 1.2, 0);
  \draw[dashed] (.3, .9) -- (.9, .9);
  \draw[dashed] (.9, .9) -- (.9, 0);
  \node at (0.9, 0) [dcirc] {};
\end{tikzpicture}
\quad + \quad
 \begin{tikzpicture}[baseline,valign]
  \draw[thick]  ( 0,  0) -- ( 1.2, 0);
  \draw[dashed] (.3, .9) -- (.3, 0);
  \draw[dashed] (.75, .9) -- (.6, 0);
  \draw[dashed] (.75, .9) -- (.9, 0);
  \node at (0.3, 0) [dcirc] {};
  \node at (0.6, 0) [dcirc] {};
  \node at (0.9, 0) [dcirc] {};
\end{tikzpicture}
\quad + \quad
 \begin{tikzpicture}[baseline,valign]
  \draw[thick]  ( 0,  0) -- ( 1.2, 0);
  \draw[dashed] (.2, .9) -- (.6, .45);
  \draw[dashed] (1., .9) to[out=-160,in=60] (.6, .45);
  \draw[dashed] (1., .9) to[out=-110,in=30] (.6, .45);
  \draw[dashed] (.6, .45) -- (.6, 0);
  \node at (0.6, 0) [dcirc] {};
  \node at (0.6, 0.45) [bcirc] {};
\end{tikzpicture}
\quad + \quad
 \begin{tikzpicture}[baseline,valign]
  \draw[thick]  ( 0,  0) -- ( 1.5, 0);
  \draw[dashed] (.3, .9) -- (.9, 0);
  \draw[dashed] (.9, .9) -- (.3, 0);
  \draw[dashed] (.9, .9) -- (1.3, 0);
  \node at (0.3, 0) [dcirc] {};
  \node at (0.9, 0) [dcirc] {};
  \node at (1.3, 0) [dcirc] {};
  \node at (0.6, 0.45) [bcirc] {};
\end{tikzpicture}
\quad + \quad
 \begin{tikzpicture}[baseline,valign]
  \draw[thick]  (-0.3,  0) -- ( 1.2, 0);
  \draw[dashed] (.6, .9) -- (.6, 0);
  \draw[dashed] (.3, 0) -- (.6, .5) -- (.9, 0);
  \draw[dashed] (.0, .9) -- (0.6, .9);
  \node at (0.3, 0.0) [dcirc] {};
  \node at (0.6, 0.0) [dcirc] {};
  \node at (0.9, 0.0) [dcirc] {};
  \node at (0.6, 0.5) [bcirc] {};
\end{tikzpicture} \notag \\[0.7em]
\quad & + \quad
\begin{tikzpicture}[baseline,valign]
  \draw[thick]  (-.3,  0) -- ( 1.2, 0);
  \draw[dashed] (.6, .9) to[out=-120,in=120] (.6, 0.3);
  \draw[dashed] (.6, .9) to[out=- 60,in= 60] (.6, 0.3);
  \draw[dashed] (.3, 0) -- (.6, .3) -- (.9, 0);
  \draw[dashed] (0, .9) -- (0, 0);
  \node at (0.0, 0.0) [dcirc] {};
  \node at (0.3, 0.0) [dcirc] {};
  \node at (0.9, 0.0) [dcirc] {};
  \node at (0.6, 0.3) [bcirc] {};
\end{tikzpicture}
\quad + \quad
 \begin{tikzpicture}[baseline,valign]
  \draw[thick]  (0,  0) -- ( 1.8, 0);
  \draw[dashed] (.6, .9) -- (.6, 0);
  \draw[dashed] (.3, 0) -- (.6, .5) -- (.9, 0);
  \draw[dashed] (1.2, 0) -- (1.35, .9) -- (1.5, 0);
  \node at (0.3, 0.0) [dcirc] {};
  \node at (0.6, 0.0) [dcirc] {};
  \node at (0.9, 0.0) [dcirc] {};
  \node at (1.2, 0.0) [dcirc] {};
  \node at (1.5, 0.0) [dcirc] {};
  \node at (0.6, 0.5) [bcirc] {};
\end{tikzpicture}
\quad + \quad
 \begin{tikzpicture}[baseline,valign]
  \draw[thick]  (-.6,  0) -- ( 1.2, 0);
  \draw[dashed] ( .4, .9) -- ( .6, .5) -- (.6, 0);
  \draw[dashed] ( .4, .9) -- (0, 0);
  \draw[dashed] ( .3, 0) -- (.6, .5) -- (.9, 0);
  \draw[dashed] (-.3, .9) -- (-.3, 0);
  \node at (-.3, 0.0) [dcirc] {};
  \node at (0.0, 0.0) [dcirc] {};
  \node at (0.3, 0.0) [dcirc] {};
  \node at (0.6, 0.0) [dcirc] {};
  \node at (0.9, 0.0) [dcirc] {};
  \node at (0.6, 0.5) [bcirc] {};
\end{tikzpicture}
\quad + \quad
\ldots
\end{align}
Notice that we have described how to evaluate all these diagrams, except for the third one, which involves the non-trivial integral
\begin{align}
  \begin{tikzpicture}[baseline,valign]
  \draw[thick]  ( 0,  0) -- ( 1.2, 0);
  \draw[dashed] (.2, .9) -- (.6, .45);
  \draw[dashed] (1., .9) to[out=-160,in=60] (.6, .45);
  \draw[dashed] (1., .9) to[out=-110,in=30] (.6, .45);
  \draw[dashed] (.6, .45) -- (.6, 0);
  \node at (0.6, 0) [dcirc] {};
  \node at (0.6, 0.45) [bcirc] {};
\end{tikzpicture}
\quad & \propto \quad
\int \frac{M^\veps \, d\tau_3 \, \, d^d x_4}
           {\big[
            (\tau_{14}^2 + \vec x_{14}^2)
            (\tau_{24}^2 + \vec x_{24}^2)^2
            (\tau_{34}^2 + \vec x_{4}^2) \big]^{1-\veps/2}}
  \notag \\[0.2em]
& = \frac{\pi^3}{|\vec x_2|^{1-\veps} x_{12}^{2-\veps}} \left(
    \frac{2}{\veps}
  + \log \frac{16 M^2 \vec x_2^2 \, x^2_{12}}
              {\tau^2_{12} + (|\vec x_1| + |\vec x_2|)^2}
  - \aleph
  + 3
+O(\veps)
\right) \, .
\label{eq:hard-int-PP2}
\end{align}
We have computed this diagram similarly to \eqref{eq:bdd-hard-int}, first bringing the integral to parametric form, using analytic regularization and keeping only terms up to $O(\veps^0)$.
The resulting parametric integrals can then be evaluated in \mathematica.\footnote{As a sanity check, we integrate the first point in \eqref{eq:hard-int-PP2} along the defect, namely $\lim_{|\vec x_1| \to 0} \int d\tau_1 \ldots \,$.
The resulting integral should be identical to the third diagram in \eqref{eq:diags-phi2}, and indeed we found perfect agreement.}

All in all, the sum of diagrams presented in the canonically normalized way reads
\begin{align}
  F_{\phi\phi^2}(r,w)
& = a_\phi a_{\phi^2}
  - \frac{3 \veps}{4} \sqrt{\frac{N+8}{2 N}} H(r,w)
  + \veps \, \frac{N+2}{\sqrt{8N(N+8)}} \,
    \xi^{-1} \log \frac{4 r}{(1 + r)^2} \notag \\
& \quad
  - \sqrt{\frac{N+8}{8 N}}
    \left[2 + \veps \log 2 - \veps \, \frac{N^2+23 N+54}{2 (N+8)^2}\right]
    \xi^{-\frac12 \Delta_{\phi^2}}
  + O(\veps^2) \, .
\label{eq:PP2-diag}
\end{align}
The one-point function $a_{\phi^2}$ is the same calculated in \eqref{eq:aP2-val}.
Note that for $N=1$ we find perfect agreement with the bootstrap correlator \eqref{eq:corr-final-pp2}.

\subsubsection{Correlator \texorpdfstring{$\vvev{\phi^2 \phi^2}$}{<phi2 phi2>}}

We conclude this section deriving the two-point function of $\phi^2$.
In this case there are ten diagrams that contribute
\begin{align}
 \vvev{ \phi^2(x) \phi^2(x) }
 \quad & = \quad
 \begin{tikzpicture}[baseline,valign]
  \draw[thick]  ( 0,  0) -- ( 1.2, 0);
  \draw[dashed] (.1, .5) to[out= 30,in= 150] (1.1, .5);
  \draw[dashed] (.1, .5) to[out=-30,in=-150] (1.1, .5);
\end{tikzpicture}
\quad + \quad
 \begin{tikzpicture}[baseline,valign]
  \draw[thick]  ( 0,  0) -- ( 1.2, 0);
  \draw[dashed] (.3, .9) -- (.9, .9);
  \draw[dashed] (.3, .9) -- (.3, 0);
  \draw[dashed] (.9, .9) -- (.9, 0);
  \node at (0.3, 0) [dcirc] {};
  \node at (0.9, 0) [dcirc] {};
\end{tikzpicture}
\quad + \quad
 \begin{tikzpicture}[baseline,valign]
  \draw[thick]  (-.3,  0) -- ( 1.2, 0);
  \draw[dashed] (0,0) -- (0.15, .9) -- (.3, 0);
  \draw[dashed] (.75, .9) -- (.6, 0);
  \draw[dashed] (.75, .9) -- (.9, 0);
  \node at (0.0, 0) [dcirc] {};
  \node at (0.3, 0) [dcirc] {};
  \node at (0.6, 0) [dcirc] {};
  \node at (0.9, 0) [dcirc] {};
\end{tikzpicture}
\quad + \quad
 \begin{tikzpicture}[baseline,valign]
  \draw[thick]  ( 0,  0) -- ( 1.2, 0);
  \draw[dashed] (.0, .5) to[out= 30,in= 150] (0.6, .5)
                         to[out= 30,in= 150] (1.2, .5);
  \draw[dashed] (.0, .5) to[out=-30,in=-150] (0.6, .5)
                         to[out=-30,in=-150] (1.2, .5);
  \node at (0.6, 0.5) [bcirc] {};
\end{tikzpicture} \notag \\[0.7em]
\quad & + \quad
 \begin{tikzpicture}[baseline,valign]
  \draw[thick]  (-.1,  0) -- ( 1.2, 0);
  \draw[dashed] (.2, 0) -- (.2, .9) -- (.6, .45);
  \draw[dashed] (1., .9) to[out=-160,in=60] (.6, .45);
  \draw[dashed] (1., .9) to[out=-110,in=30] (.6, .45);
  \draw[dashed] (.6, .45) -- (.6, 0);
  \node at (0.2, 0) [dcirc] {};
  \node at (0.6, 0) [dcirc] {};
  \node at (0.6, 0.45) [bcirc] {};
\end{tikzpicture}
\quad + \quad
 \begin{tikzpicture}[baseline,valign]
  \draw[thick]  ( 0,  0) -- ( 1.2, 0);
  \draw[dashed] (.3, .9) -- (.9, .9);
  \draw[dashed] (.3, .9) -- (.9, 0);
  \draw[dashed] (.9, .9) -- (.3, 0);
  \node at (0.3, 0) [dcirc] {};
  \node at (0.9, 0) [dcirc] {};
  \node at (0.6, 0.45) [bcirc] {};
\end{tikzpicture}
\quad + \quad
 \begin{tikzpicture}[baseline,valign]
  \draw[thick]  ( 0,  0) -- ( 1.5, 0);
  \draw[dashed] (.6, .9) -- (1.2, .9);
  \draw[dashed] (.6, .9) -- (.6, 0);
  \draw[dashed] (.3, 0) -- (.6, .5) -- (.9, 0);
  \draw[dashed] (1.2, .9) -- (1.2, 0);
  \node at (0.3, 0.0) [dcirc] {};
  \node at (0.6, 0.0) [dcirc] {};
  \node at (0.9, 0.0) [dcirc] {};
  \node at (1.2, 0.0) [dcirc] {};
  \node at (0.6, 0.5) [bcirc] {};
\end{tikzpicture}
\quad + \quad
\begin{tikzpicture}[baseline,valign]
  \draw[thick]  (-.6,  0) -- ( 1.2, 0);
  \draw[dashed] (.6, .9) to[out=-120,in=120] (.6, 0.3);
  \draw[dashed] (.6, .9) to[out=- 60,in= 60] (.6, 0.3);
  \draw[dashed] (.3, 0) -- (.6, .3) -- (.9, 0);
  \draw[dashed] (-.3, 0) -- (-.15, .9) -- (0, 0);
  \node at (-.3, 0.0) [dcirc] {};
  \node at (0.0, 0.0) [dcirc] {};
  \node at (0.3, 0.0) [dcirc] {};
  \node at (0.9, 0.0) [dcirc] {};
  \node at (0.6, 0.3) [bcirc] {};
\end{tikzpicture}\notag \\[0.7em]
\quad & + \quad
 \begin{tikzpicture}[baseline,valign]
  \draw[thick]  (-.2,  0) -- ( 1.4, 0);
  \draw[dashed] (0, 0) -- (.3, .9) -- (.9, 0);
  \draw[dashed] (1.2, 0) -- (.9, .9) -- (.3, 0);
  \node at (0.0, 0) [dcirc] {};
  \node at (0.3, 0) [dcirc] {};
  \node at (0.9, 0) [dcirc] {};
  \node at (1.2, 0) [dcirc] {};
  \node at (0.6, 0.45) [bcirc] {};
\end{tikzpicture}
\quad + \quad
 \begin{tikzpicture}[baseline,valign]
  \draw[thick]  (-.9,  0) -- ( 1.2, 0);
  \draw[dashed] ( .4, .9) -- ( .6, .5) -- (.6, 0);
  \draw[dashed] ( .4, .9) -- (0, 0);
  \draw[dashed] ( .3, 0) -- (.6, .5) -- (.9, 0);
  \draw[dashed] (-.6, .0) -- (-.45, .9) -- (-.3, 0);
  \node at (-.6, 0.0) [dcirc] {};
  \node at (-.3, 0.0) [dcirc] {};
  \node at (0.0, 0.0) [dcirc] {};
  \node at (0.3, 0.0) [dcirc] {};
  \node at (0.6, 0.0) [dcirc] {};
  \node at (0.9, 0.0) [dcirc] {};
  \node at (0.6, 0.5) [bcirc] {};
\end{tikzpicture}
\quad + \quad
\ldots
\end{align}
However, all of them are equivalent to the diagrams previously studied.
Carefully combining the results gives
\begin{align}
  F_{\phi^2\phi^2}(r,w)
& = a^2_{\phi^2}
  + \xi^{-\Delta_{\phi^2}}
  - \veps \, \frac{N+2}{2 N} \, \xi^{-1}
    \log \frac{4 r}{\left(1+r\right)^2} \notag \\
& \quad
  + \frac{N+8}{2 N}
    \left(1 + \veps \log 2 - \veps \, \frac{N^2+23 N+54}{2 (N+8)^2} \right)
    \xi^{\Dp-\Dpp}
    \notag  \\
& \quad
  + \veps \left(
     \frac{3 N + 24}{8 N}
     + \frac{N+2}{2 N} \, \xi^{-1}
    \right) H(r,w)
  + O(\veps^2) \, .
  \label{eq:P2P2-diag}
\end{align}
As a final comment, observe that for $N > 1$ both $F_{\phi\phi^2}$ and $F_{\phi^2\phi^2}$ would be difficult to compute using bootstrap methods, since one needs to first solve a mixing problem.
Therefore, from the bootstrap point of view it is surprising that the two-point functions \eqref{eq:PP-diag}, \eqref{eq:PP2-diag} and \eqref{eq:P2P2-diag} have such a similar structure, and it would be interesting to understand if there is a CFT argument for why this should be the case.

\section{OPE analysis}
\label{sec:ope-analysis}

In section \ref{sec:feyndiag} we determined two-point functions in the presence of a localized magnetic field line defect in the $\veps$--expansion.
These correlation functions contain CFT data for several infinite families of operators, as we discuss below.
However, to extract the physically interesting one-point coefficients $a_\Om$, we need to remove the an extra three-point coefficient $\lambda_{12\Om}$.
For this reason, we start in section \ref{sec:review-WF} with a review of the Wilson-Fisher fixed point without a defect.
We are then ready to present bulk data in section \ref{sec:bulk-ope} and defect data in section \ref{sec:defect-ope}, paying particular attention to families of non-degenerate operators.

\subsection{Review of Wilson-Fisher CFT data}
\label{sec:review-WF}

We start reviewing the lowest-twist trajectories for the Wilson-Fisher $O(N)$ model, but we refer to \cite{Henriksson:2022rnm} for a more thorough discussion.
The fundamental field $\phi_a(x)$, which transforms as a vector under $O(N)$, has the same dimension as a free field at the order we are working
\begin{align}
 \Dp = \frac{d-2}{2} + O(\veps^2) \, .
\end{align}
We are interested on the operators that appear in the OPE of $\phi_a \times \phi_b$ that transform as singlets or rank-two symmetric-traceless tensors.
The simplest operators are $\phi_c^2$ and $T_{ab} = \phi_a \phi_b - \frac{\delta_{ab}}{N} \phi_c^2$, whose dimension and OPE coefficients read
\begin{align}
 \Delta_{\phi^2}
&= 2 \Delta_\phi + \gpp \veps + O(\veps^2) \, , \qquad
 \lambda_{\phi\phi\phi^2}^2
 = \frac{2}{N} (1 - \gpp \veps ) + O(\veps^2) \, , \qquad
 \gpp = \frac{N+2}{N+8} \, , \label{eq:diml-p2} \\
 \Delta_{T}
&= 2 \Delta_\phi + \gamma_{T} \veps + O(\veps^2) \, , \qquad \;\;
 \lambda_{\phi\phi T}^2
 = 2 (1 - \gamma_T \veps ) + O(\veps^2) \, , \qquad  \quad \,
 \gamma_{T} = \frac{2}{N+8} \, . \label{eq:diml-Tab}
\end{align}
More generally we can construct singlet and symmetric-traceless operators of approximate twist two, that schematically read
\begin{align}
 \Om^S_{\ell,0}
 \sim \partial_{\mu_1} \ldots \partial_{\mu_\ell} \phi^2_a \, , \qquad
 \Om^T_{\ell,0}
 \sim \partial_{\mu_1} \ldots \partial_{\mu_\ell}
 \left( \phi_a \phi_b - \frac{\delta_{ab}}{N} \phi_c^2 \right) \, .
\end{align}
These operators correspond to higher-spin currents in the free theory.
At the order we work, their dimension remains protected and their CFT data reads as in free theory:
\begin{align}
 & \Delta_{\Om^S_{\ell,0}}
 = 2 \Dp + \ell + O(\veps^2) \, , \qquad
 \lambda_{\phi\phi\Om^S_{\ell,0}}^2
 = \frac{2^{\ell+1} (\Dp)_\ell^2}{N \ell! (2 \Dp+\ell-1)_\ell}
 + O(\veps^2) \, , \\
 & \Delta_{\Om^T_{\ell,0}}
 = 2 \Dp + \ell + O(\veps^2) \, , \qquad
 \lambda_{\phi\phi\Om^T_{\ell,0}}^2
 = \frac{2^{\ell+1} (\Dp)_\ell^2}{\ell! (2 \Dp+\ell-1)_\ell}
 + O(\veps^2) \, .
\end{align}
We can also construct higher-twist families, for example at $n=1$
\begin{align}
 \Om^S_{\ell,1}
 \; \sim \;
 \partial^2 \partial_{\mu_1} \ldots \partial_{\mu_\ell} \phi^2_a \, ,
 \;\;
 \partial_{\mu_1} \ldots \partial_{\mu_\ell} (\phi^2_a)^2 \, .
\end{align}
Here we wrote two schematic operators with the same quantum numbers.
This follows because $\phi$ has dimension one in the free theory limit, so the Laplacian $\partial^2$ has the same dimension as $\phi_a^2$.
All in all, at each spin there can be multiple inequivalent operators $\Om_i$ with nearly degenerate scaling dimensions.
In an OPE decomposition one can only extract the sum over this nearly degenerate contributions, which we denote by $\vvev{ \lambda_{\phi\phi\Om}^2 } = \sum_i \lambda_{\phi\phi\Om_i}^2$.
For the case of present interest, the OPE coefficients were found in \cite{Alday:2017zzv,Henriksson:2018myn}
\begin{align}
  \vvev{ \lambda_{\phi\phi \Om^S_{\ell,1}}^2 }
  & = \frac{2^\ell \Gamma (\ell+2)^2}{\Gamma (2 \ell+3)}
   \, \frac{(N+2)\left(\ell^2+3 \ell+8\right)}{4 N (N+8)^2 (\ell+1) (\ell+2)}
   \, \veps^2 + O(\veps^3) \, , \\
  \vvev{ \lambda_{\phi\phi \Om^T_{\ell,1}}^2 }
  & = \frac{2^\ell \Gamma (\ell+2)^2}{\Gamma (2 \ell+3)}
   \, \frac{(N+2)\left(\ell^2+3 \ell+8\right) - 4N}{4 N (N+8)^2 (\ell+1) (\ell+2)}
   \, \veps^2 + O(\veps^3) \, .
\end{align}
Finally, let us mention that higher-twist families are suppressed in $\veps$, as was first observed in \cite{Liendo:2012hy}, and later proved with analytic bootstrap in \cite{Alday:2017zzv,Henriksson:2018myn}:
\begin{align}
 \vvev{ \lambda_{\phi\phi \Om^{S/T}_{\ell,n\ge2}}^2 }
 & = O(\veps^3) \, .
\end{align}
Furthermore, below we need the coupling between $\phi^2$ and the leading singlet trajectory \cite{Bertucci:2022ptt}:
\begin{align}
 \lambda_{\phi^2\phi^2\phi^2}
&= \frac{2 \sqrt{2}}{\sqrt{N}} \left(
     1
    - \frac{3 (N+2)}{2 (N+8)} \, \veps
    + O(\veps^2)
 \right) \, , \\
 \lambda_{\phi^2\phi^2\Om_{\ell,0}^S}
&= 2 \lambda_{\phi\phi\Om_{\ell,0}^S} \left(
   1
  + \frac{N+2}{N+8} \left(H_\ell-1\right) \veps
  + O(\veps)
  \right) \, .
\end{align}

Turning our attention to the $\phi \times \phi^2$ OPE, we find operators $\Om_{\ell,n}^V$ that are vectors under $O(N)$.
These operators are highly degenerate, and even the leading family has mixing.
For this reason, we restrict our attention to the Ising model $N=1$, where the situation improves.
Let us look at triple-twist operators
\begin{align}
 \Om^{-}_{\ell,0,i}
 \sim \partial_{\mu_1} \ldots \partial_{\mu_\ell} \phi^3 \, ,
 \label{eq:defOmi}
\end{align}
where $i = 1,...,d_\ell$ distinguishes the $d_\ell$ degenerate operators, and $\Om^{-}$ indicates these are odd under $\mathbb{Z}_2$.
It was shown in \cite{Kehrein:1992fn} that at each spin only one degenerate operator acquires anomalous dimension up to $O(\veps^2)$ corrections.
Furthermore, this operator is the only that has non-vanishing OPE coefficient \cite{Bertucci:2022ptt} at the same order.
We call this operator $\Om^{-}_{\ell,0} \equiv \Om^{-}_{\ell,0,1}$, and restrict our attention to it.
Its CFT data reads \cite{Bertucci:2022ptt}
\begin{align}
 \Delta_{\Om^{-}_{\ell,0}}
&= \Dp
 + \Delta_{\phi^2}
 + \frac23 \frac{(-1)^\ell}{\ell+1} \, \veps
 + O(\veps^2) \, , \\[0.2em]
 \lambda^2_{\phi\phi^2\Om^{-}_{\ell,0}}
&= \frac{2^{\ell+1} \Gamma(\ell+2)^2(\ell+1+2(-1)^\ell)}{\Gamma(2\ell+3)}
 + O(\veps) \, .
\end{align}
The fact that $\Om^{-}_{\ell,0,i\ge2}$ have subleading OPE coefficients and anomalous dimensions is what allowed us to treat the triple-twist family as non-degenerate in the bootstrap analysis of section \ref{sec:bootstrap}.
For the $O(N)$ model for $N>1$ this is no longer the case, so the bootstrap calculation would be more complicated, and also we cannot extract unmixed CFT data for these operators below.

\subsection{Bulk CFT data}
\label{sec:bulk-ope}

\subsubsection{Singlet channel}

Let us start presenting the CFT data in the $O(N)$-singlet sector, which is probed by the $F_{\phi\phi}$ and $F_{\phi^2\phi^2}$ correlators.
Remember that $F_{\phi\phi}$ decomposes into two symmetry channels \eqref{eq:blk-tens-struct}, the singlet one being
\begin{align}
  F_S(r,w)
  = \xi^{-\Dp}
  + \frac{a_\phi^2}{N}
  + \frac{N+2}{4 N} \, \veps \, H(r,w)
  + O(\veps^2) \, .
\end{align}
We can decompose this correlator in conformal blocks, and the result captures contributions of twist-two and twist-four families
\begin{align}
  \xi^{\Delta_\phi} F_S(r,w)
& = 1
    + \lambda_{\phi\phi\phi^2} a_{\phi^2} f_{\Delta_{\phi^2},0} 
    + \sum_{\substack{\ell=2 \\ \ell \text{ even}}}^\infty 2^{-\ell}
    \lambda_{\phi\phi\Om_{\ell,0}^S}
    a_{\Om_{\ell,0}^S}
    f_{2\Delta_\phi+\ell,\ell} 
    \notag \\
& \quad
    + \sum_{\substack{\ell=0 \\ \ell \text{ even}}}^\infty 2^{-\ell}
    \vvev{ \lambda_{\phi\phi\Om_{\ell,1}^S}
    a_{\Om_{\ell,1}^S} }
    f_{2\Delta_\phi+\ell+2,\ell} 
    + O(\veps^2) \, .
\end{align}
The precise values of the CFT data can be obtained from \eqref{eq:exp-H-PP} and \eqref{eq:exp-bid-PP}.
In particular, it can be checked that the expansion reproduces $a_{\phi^2}$ as reported in \eqref{eq:aP2-val}.
Similarly, we can extract the one-point functions of the twist-two family, which are new predictions
\begin{align}
  a_{\Om_{\ell,0}^S}
& = \frac{(N+8)}{\sqrt{N}}
    \frac{(\ell!)^3}{2^{\frac{3\ell+5}{2}} \left(\frac{\ell}{2}!\right)^4 \sqrt{(2 \ell)!}}
    \bigg[
     1 \notag \\
&    \qquad
     +\frac{\veps}{2} \left(
        \frac{N^2-3N-22}{(N+8)^2}
        -2 H_{\frac{\ell-1}{2}}+H_{\ell-\frac{1}{2}}+2 H_\ell-H_{2 \ell}
      \right)+O(\veps^2)
    \bigg] \, .
 \label{eq:aOS}
\end{align}
Moving to twist four presents the complication of mixing between nearly-degenerate operators, so we can only extract the average CFT data:
\begin{align}
  \vvev{ a_{\Om_{\ell,1}^S} }
& \equiv
    \frac{\vvev{\lambda_{\phi\phi\Om_{\ell,1}^S} a_{\Om_{\ell,1}^S}}}
         {\vvev{\lambda^2_{\phi\phi\Om_{\ell,1}^S}}^{1/2}} \notag \\
& = \frac{2^{-\frac{3\ell+6}{2}} ((\ell+1)!)^3}
         {\left(\frac{\ell}{2}!\right)^4 \sqrt{(2 \ell+2)!}}
    \frac{(N+8)^2}{\sqrt{N(N+2)} \sqrt{(\ell+1) (\ell+2) (\ell^2+3\ell+8)}}
  + O(\veps) \, ,
\end{align}
For non-degenerate operators, the average has to agree with the actual value of $a_\Om$. For instance, the operator $\Om_{0,1}^S = \phi^4$ is non-degenerate, and indeed we find agreement with \eqref{eq:aP4}.

The singlet channel is also probed by the $F_{\phi^2\phi^2}$ correlation function.
In this case, we gain access to more high-twist families, although these are in general degenerate:\footnote{The notation $\vvev{ \lambda_\Om a_\Om f_{\Delta_\Om,\ell}}$ expresses the fact that we can extract $\vvev{ \lambda_\Om a_\Om}$ and $\vvev{ \lambda_\Om a_\Om \gamma_\Om}$, so the anomalous dimension is mixed with the OPE coefficients. }
\begin{align}
  \xi^{\Dpp} F_{\phi^2\phi^2}(r,w)
& = 1
    + \lambda_{\phi^2\phi^2\phi^2} a_{\phi^2} f_{\Delta_{\phi^2},0}
    + \sum_{\substack{\ell=2 \\ \ell \text{ even}}}^\infty 2^{-\ell}
    \lambda_{\phi^2\phi^2\Om_{\ell,0}^S}
    a_{\Om_{\ell,0}^S}
    f_{2\Delta_\phi+\ell,\ell} \notag \\
& \quad
    + \sum_{n=1}^\infty
      \sum_{\substack{\ell=0 \\ \ell \text{ even}}}^\infty 2^{-\ell}
    \vvev{ \lambda_{\phi^2\phi^2\Om_{\ell,n}^S}
    a_{\Om_{\ell,n}^S}
    f_{\Delta_{\Om_{\ell,n}^S},\ell} }
    + O(\veps^2) \, .
\end{align}
We have checked that the CFT data of $\phi^2$ and the twist-two family agree with \eqref{eq:aOS}.
It is not hard to extract the averaged CFT data from the formulas in appendix \ref{sec:block-exp}, but the results are not particularly illuminating.
Instead, we content ourselves by mentioning that we checked once again that the one-point function $a_{\phi^4}$ is correctly reproduced by our results.

\subsubsection{Symmetric-traceless channel}

We can similarly analyse symmetric-traceless operators.
In this case, the relevant part of $F_{\phi\phi}$ is
\begin{align}
  F_T(r,w)
  = a_\phi^2
  + \frac{\veps}{2} \, H(r,w) \, .
\end{align}
The structure of the block expansion is identical to the singlet channel, except the bulk identity is not present:
\begin{align}
  \xi^{\Delta_\phi} F_T(r,w)
& = \lambda_{\phi\phi T} a_{T} f_{\Delta_T,0}
    + \sum_{\substack{\ell=2 \\ \ell \text{ even}}}^\infty 2^{-\ell}
    \lambda_{\phi\phi\Om_{\ell,0}^T}
    a_{\Om_{\ell,0}^T}
    f_{2\Delta_\phi+\ell,\ell} \notag \\
& \quad
    + \sum_{\substack{\ell=0 \\ \ell \text{ even}}}^\infty 2^{-\ell}
    \vvev{ \lambda_{\phi\phi\Om_{\ell,1}^T}
    a_{\Om_{\ell,1}^T} }
    f_{2\Delta_\phi+\ell+2,\ell}
    + O(\veps^2) \, .
\end{align}
We have checked that this expansion leads to the correct value of $a_T$ in \eqref{eq:aT-val}.
The one-point functions of the twist-two family are again new predictions:
\begin{align}
 a_{\Om_{\ell,0}^T}
 = a_{\Om_{\ell,0}^S} \sqrt{N} \, .
\end{align}
Finally, we have degenerate twist-four operators, that have the following averaged CFT data
\begin{align}
  \vvev{ a_{\Om_{\ell,1}^T} }
& = \frac{2^{-\frac{3\ell+6}{2}} ((\ell+1)!)^3}
         {\left(\frac{\ell}{2}!\right)^4 \sqrt{(2 \ell+2)!}}
    \frac{(N+8)^2}
         {\sqrt{(\ell+1) (\ell+2) [(\ell^2+3 \ell+8) (N+2)-4 N ]}}
  + O(\veps) \, .
\end{align}
Once again, the scalar operator $\ell =0$ agrees with the prediction for $a_{\phi^2 T}$ in \eqref{eq:aP4}.

\subsubsection{Vector channel}

Lastly, we consider the CFT data of vector operators under $O(N)$, which can be accessed from $F_{\phi\phi^2}$:
\begin{align}
  \xi^{\frac12(\Delta_\phi+\Delta_{\phi^2})} F_{\phi\phi^2}(r,w)
& = \lambda_{\phi\phi\phi^2} a_{\phi} f_{\Dp,0}
    + \sum_{\substack{\ell=2 \\ \ell \text{ even}}}^\infty 2^{-\ell}
    \vvev{
    \lambda_{\phi\phi^2\Om_{\ell,0}^V}
    a_{\Om_{\ell,0}^V}
    f_{\Delta_{\Om_{\ell,0}^V},\ell} } \notag \\
& \quad
    + \sum_{n=1}^\infty
      \sum_{\substack{\ell=0 \\ \ell \text{ even}}}^\infty 2^{-\ell}
    \vvev{ \lambda_{\phi\phi\Om_{\ell,n}^V}
    a_{\Om_{\ell,n}^V} }
    f_{2\Delta_\phi+\ell+2n,\ell}
    + O(\veps^2) \, .
\end{align}
First, we checked that the operator $\phi$ appears with the correct one-point function given in \eqref{eq:val-aP}.
Second, we checked that the scalar in the leading-twist family, which should be identified as $\Om_{0,0}^V \sim \phi^2 \phi_a$, decouples from the expansion.
This is expected because of multiplet recombination, since $\partial^2 \phi \propto \veps \phi^2 \phi_a$ in the interacting theory, so $\phi^2 \phi_a$ becomes a conformal descendant.
However, we show in appendix \ref{sec:block-exp} that this expansion has to be done with great care.

Finally, we remind that the triple-twist family is degenerate \cite{Kehrein:1992fn}.
When $N=1$ we can ignore this degeneracy at tree-level, as discussed below equation \eqref{eq:defOmi}, and we extract
\begin{align}
 a_{\Om_{\ell,0}^-}
 = -\frac{27 \, \Gamma (2 \ell+3)^{1/2} }
         {2^{\frac{3 \ell}{2}+4}  \sqrt{\ell+3} \, \Gamma (\ell+2)}
 + O(\veps) \, .
\end{align}
With the methods in appendix \ref{sec:block-exp}, we can also extract the $O(\veps)$ correction, but it is not guaranteed it corresponds to unmixed CFT data, so we ignore it here.
Similarly, with the help of the appendix it is possible to extract averaged CFT data for $N > 1$ and for the higher-twist families $n \ge 1$.

\subsection{Defect CFT data}
\label{sec:defect-ope}

We now move on to the defect conformal block expansion.
In this case, the correlators probe operators $\wh \Om_{s,n}^S$ and $\wh \Om_{s,n}^V$ that are singlet or vector under the $O(N-1)$ symmetry preserved by the defect.
These defect operators have tree-level dimension
\begin{align}
 \Delta_{\Om_{s,n}^S} \approx \Dp + s + n \, , \qquad
 \Delta_{\Om_{s,n}^V} \approx \Dp + s + n \, .
\end{align}
As in the bulk, there is a large degeneracy of defect operators in each sector, except for the leading trajectory which is non-degenerate
\begin{align}
 \wh \Om_{s,0}^S
 \sim \partial_{i_1} \ldots \partial_{i_s} \phi_1 \, , \qquad
 \wh \Om_{s,0}^V
 \sim \partial_{i_1} \ldots \partial_{i_s} \phi_{\hat a} \, .
\end{align}
Above $i = 1, \ldots, d-1$ labels directions orthogonal to the defect and $\hat a = 2, \ldots, N$ labels $O(N-1)$ indices.
It is easy to see that already at $n=1$ there is at least two-fold degeneracy, namely
\begin{align}
 \wh \Om_{s,1}^S
 \; \sim \;
 \partial_{i_1} \ldots \partial_{i_s} \phi^2_1 \, , \;\;
 \partial_{i_1} \ldots \partial_{i_s} \phi^2_{\hat a} \, , \; \ldots
\end{align}
The mixing problem for $n=1$ and $s = 0$ has been solved in \cite{Cuomo:2021kfm,Gimenez-Grau:2022czc} using diagrammatic techniques.
Below we shall not attempt to solve mixing problems, and instead the focus is on the leading-twist family.

\subsubsection{Singlet channel}

Let us start analyzing singlet operators under $O(N-1)$, which can be accessed from the three correlators at our disposal.
Considering $F_{\phi\phi}$, we need to take the contribution from the correct tensor structure \eqref{eq:def-tens-struct}:
\begin{align}
 F_{\wh S}(r,w)
 = a_\phi^2
 + \xi^{-\Dp}
 + \frac{3\veps}{4} \, H(r,w) \, .
\end{align}
We then perform the defect conformal block expansion, and observe that at this order in $\veps$ only the leading-twist trajectory appears
\begin{align}
 F_{\wh S}(r,w)
 = a_\phi^2
 + \sum_{s=0}^\infty 2^{-s} b^2_{\phi \wh \Om^S_{s,0}}
   \wh f_{\wh\Delta_{\wh \Om^S_{s,0}},s}
 + O(\veps^2) \, .
\end{align}
The algorithm to obtain the OPE coefficients is detailed in appendix \ref{sec:app-def-exp}, and here we simply report the CFT data:\footnote{When this work was being prepared for publication, reference \cite{Giombi:2022vnz} appeared. The anomalous dimensions in \eqref{eq:DhOs} and \eqref{eq:dimVh} agree perfectly with the results in their appendix C.}
\begin{align}
 & \wh\Delta_{\wh \Om^S_{s,0}}
 = 1
 + s
 - \frac{s-1}{2 s+1} \veps
 + O(\veps^2) \, , \label{eq:DhOs}\\
 & b_{\phi \wh \Om^S_{s,0}}
 = 2^{s/2} \left(
   1
 - \frac{3 \veps}{2 (2 s+1)^2}
 - \frac{2 (s-1) H_s + 3 H_{s-\frac{1}{2}}}{4 (2 s+1)} \, \veps
 + O(\veps^2)
 \right) \, . \label{eq:bPOs}
\end{align}
We can perform several consistency checks on this result.
First, focusing on $s = 0$ we recover the correct dimension of the operator $\wh \Om^S_{0,0} = \wh\phi$, as well as $b_{\phi\wh\phi}$ computed in \eqref{eq:b-diag}.
Second, we consider the displacement operator $\wh \Om^S_{1,0} = \vec D \sim \vec\partial \phi$, for which the scaling dimension agrees with the Ward identity prediction $\wh \Delta_D = 2$.

We can also access the singlet channel using the mixed two-point function between $\phi_a$ and $\phi^2$ given in \eqref{eq:PP2-diag}.
In this case, both the leading-twist and higher even-twist families appear
\begin{align}
 F_{\phi\phi^2}(r,w)
&= a_\phi a_{\phi^2}
 + \sum_{s=0}^\infty 2^{-s}
   b_{\phi \wh \Om^S_{s,0}} b_{\phi^2 \wh \Om^S_{s,0}}
   \wh f_{\wh\Delta_{\wh \Om^S_{s,0}},s} \notag \\
& \quad
 + \sum_{n=0}^\infty \sum_{s=0}^\infty 2^{-s}
   \vvev{ b_{\phi \wh \Om^S_{s,2n+1}} b_{\phi^2 \wh \Om^S_{s,2n+1}} }
   \wh f_{2+s+2n,s}
 + O(\veps^2) \, .
\end{align}
For the leading-twist family, we obtain once again the scaling dimensions in \eqref{eq:DhOs}.
Regarding OPE coefficients, we use \eqref{eq:bPOs} to disentangle their product:
\begin{align}
 b_{\phi^2 \wh \Om^S_{s,0}}
&= - 2^{\frac{s-1}{2}} \sqrt{\frac{N+8}{N}} \Bigg(
   1
 + \frac{N-4}{4 (N+8)} \left(H_s-2 \log 2\right) \veps
 + \frac{3 H_s - 3 H_{s-\frac{1}{2}}}{4 (2 s+1)}  \, \veps \notag \\
& \hspace{12em}
 - \frac{N^2+23 N+54}{4(N+8)^2}  \, \veps
 - \frac{3\veps}{2(2 s+1)^2}
 + O(\veps^2)
 \Bigg) \, .
 \label{eq:bP2OS}
\end{align}
Here we have performed yet another consistency check, by seeing that the Ward identity \eqref{eq:wi} is satisfied for arbitrary $N$.
As a curiosity, note that the even-twist CFT data admits a simple closed form expression:
\begin{align}
 \vvev{ b_{\phi \wh \Om^S_{s,2n+1}} b_{\phi^2 \wh \Om^S_{s,2n+1}} }
 = - \frac{N+2}{\sqrt{N (N+8)}}
   \frac{2^{\frac{2s-1}{2}} (s+1)_{2 m+1}}
        {(2 m+1) (m+s+1) \left(s+\frac{3}{2}\right)_{2 m}} \, .
\end{align}

Finally we look at the two-point function of $\phi^2$, which admits the expansion
\begin{align}
 F_{\phi^2\phi^2}(r,w)
 = a_{\phi^2}^2
 + \sum_{s=0}^\infty 2^{-s} \,
   b_{\phi^2 \wh \Om^S_{s,0}}^2
   \wh f_{\wh\Delta_{\wh \Om^S_{s,0}},s}
 + \sum_{n=1}^\infty \sum_{s=0}^\infty 2^{-s} \,
   \vvev{ b_{\phi^2 \wh \Om^S_{s,n}}^2
   \wh f_{\wh\Delta_{\wh \Om^S_{s,n}},s} }
 + O(\veps^2) \, .
\end{align}
We have checked that the leading-twist expansion agrees with \eqref{eq:DhOs} and \eqref{eq:bP2OS}.
The CFT data for higher-twists follows from the methods of appendix \ref{sec:app-def-exp}, but the result is not particularly illuminating.

\subsubsection{Vector channel}

Finally, we turn our attention to operators which are vector under $O(N-1)$, which can only be obtained from the two-point function of $\phi_a$ in \eqref{eq:PP-diag}:
\begin{align}
 F_{\wh V}(r,w)
 = \xi^{-\Dp}
 + \frac{\veps}{4} \, H(r,w)
 + O(\veps^2) \, .
\end{align}
The structure of the CFT data is similar as for $F_{\wh S}$, with the only difference that the defect identity is not present (as expected):
\begin{align}
 F_{\wh V}(r,w)
 = \sum_{s=0}^\infty 2^{-s} b^2_{\phi \wh \Om^V_{s,0}}
   \wh f_{\wh\Delta_{\wh \Om^V_{s,0}},s}
 + O(\veps^2) \, .
\end{align}
Using the methods from the appendix we extract the CFT data
\begin{align}
 & \wh\Delta_{\wh \Om^V_{s,0}}
 = 1
 + s
 - \frac{s}{2 s+1} \veps
 + O(\veps^2) \, , \label{eq:dimVh} \\
 & b_{\phi \wh \Om^V_{s,0}}
 = 2^{s/2} \left(
   1
 - \frac{\veps}{2 (2 s+1)^2}
 - \frac{2 s H_s+H_{s-\frac{1}{2}}}{4 (2 s+1)} \, \veps
 + O(\veps^2)
 \right) \, .
\end{align}
The tilt operator $t$ corresponds to $s=0$, and indeed we recover the correct dimension $\wh\Delta_t = 1$ and the correct coefficient $b_{\phi t}$ in \eqref{eq:b-diag}.

\section{Conclusions}
\label{sec:conclusions}

The main result of this paper are the mixed two-point functions of the order parameter $\phi$ and energy operator $\phi^2$ in the presence of a localized magnetic field line defect \cite{Allais:2014fqa,2014arXiv1412:3449A,2017PhRvB:95a4401P,Cuomo:2021kfm}, see equations \eqref{eq:PP-diag}, \eqref{eq:PP2-diag} and \eqref{eq:P2P2-diag}.
We started using analytic bootstrap for these mixed correlators, focusing mostly on the $N=1$ Wilson-Fisher fixed point.
Having access to mixed correlators allowed us to impose multiple consistency conditions that fixed uniquely the final result.
For $N>1$ we bootstrapped a single correlator, and although the result has one unfixed parameter, we believe the extension to mixed correlators would also fix this parameter.
We have supplemented the bootstrap analysis with perturbative calculations using Feynman diagrams, finding perfect agreement when comparison was possible.
Finally, we performed a detailed OPE analysis that provides new predictions for CFT data that might be useful for future studies.

This work demonstrates that the bootstrap is a powerful method to study conformal defects.
Although our analysis relies on the $\veps$--expansion, the analytic techniques we used hold non-perturbatively, so the natural next step is to go beyond pertubation theory.
One idea is to consider truncated crossing equations \cite{Gliozzi:2013ysa,Gliozzi:2014jsa}, an approach that works well for boundary CFT \cite{Gliozzi:2015qsa,Padayasi:2021sik}.
Although preliminary results \cite{truncatedNotes} are still somewhat far from perturbative estimates \cite{Cuomo:2021kfm} and Monte Carlo simulations \cite{2017PhRvB:95a4401P,2014arXiv1412:3449A}, perhaps the method of \cite{Laio:2022ayq} to include more operators in the truncation can help.
An alternative is to use analytic bootstrap technology, similar in spirit to \cite{Simmons-Duffin:2016wlq,Albayrak:2019gnz,Liu:2020tpf,Caron-Huot:2020ouj,Atanasov:2022bpi}, but in a setup with a line defect.
Note that defect CFT kinematics are somewhat similar to CFT at finite temperature, where this approach was successful \cite{Iliesiu:2018zlz}.
To pursue either of these directions it will be important to remember the lessons from the present work.
Namely, that to uniquely identify the localized magnetic line defect it is a good idea to bootstrap mixed correlators and to impose the Ward identities satisfied by the displacement operator.

If we allow ourselves to remain in the perturbative regime, there are many models that can be studied with similar techniques to ours.
One natural candidate is the ``magnetic Wilson line'' \cite{2000cond.mat..4156S,2000PhRvB..6115152V} recently revisited in \cite{Cuomo:2022xgw}, for which we are planning a more detailed $\veps$--expansion analysis.
Similarly, the study of bulk two-point functions might provide useful information about line defects in the Gross-Neveau model \cite{Giombi:2021cnr,Giombi:2022vnz} or in melonic CFTs \cite{Popov:2022nfq}.
In the context of supersymmetric theories with a holographic description, there are many interesting defects which are engineered by intersecting branes in the string dual.
The example of a Maldacena-Wilson line in $\Nm=4$ SYM at strong coupling has been studied with bootstrap techniques in \cite{Barrat:2021yvp,Barrat:2022psm}.
In principle, these results can be used to obtain the leading stringy corrections to the effective action of the dual branes \cite{wittenDiagrams}.\footnote{I thank Victor Rodriguez and Yifan Wang for useful discussions on this point.}
Similar remarks also apply to M2-brane defects in $6d$ $(2,0)$ theory, as well as the half-BPS line defect in $3d$ $\Nm=8$ ABJM theory.

Finally, it would also be interesting to clarify several technical aspects of our calculation.
The first is the assumption about Regge behavior, namely that the correlator is constant as $|w| \to \infty$.
At the moment we lack a non-perturbative bound on Regge behavior for defect CFT, except for the case of boundary CFT \cite{Kaviraj:2018tfd,Mazac:2018biw}, and it would be interesting to see if such a bound could be derived.
Relatedly, from the AdS/DCFT perspective \cite{Kaviraj:2018tfd,Rastelli:2017ecj,Mazac:2018biw} one can add an arbitrary number of Witten diagrams to a correlator while preserving crossing, but worsening Regge behavior as much as desired.
It would be interesting to understand why in theories of interest these contributions seem to be absent.
One last technical problem we should mention are the families of nearly-degenerate operators, which pose the main obstacle to pushing analytic bootstrap to higher perturbative orders.
It would be fascinating if the bootstrap of higher-point correlators \cite{Bercini:2020msp,Antunes:2021kmm} gives a handle on these mixing problems.

\section*{Acknowledgements}

I am particularly grateful to Gabriel Cuomo, Zohar Komargodski and Avia Raviv-Moshe for discussions that inspired this paper, and to Pedro Liendo for comments on the draft.
I am also thankful to Julien Barrat, Edo Lauria, Pedro Liendo and Philine van Vliet for collaboration on related projects, and to Johan Henriksson for useful correspondence.
Preliminary versions of this work have been presented in Paris and Pisa, and I would like to thank these groups for many interesting comments.
During most of this work, I have been supported by the DFG through the Emmy Noether research group ``The Conformal Bootstrap Program'' project number 400570283, while I am now supported by a Simons fellowship at IHES.
I would also like to thank EPFL for hospitality during part of this work.

\appendix

\section{Review of defect CFT}
\label{sec:review-dcft}

\subsection{Kinematics}

In defect conformal field theory (DCFT) one considers two theories, the bulk and the defect CFTs, which interact in a non-trivial fashion.
On the one hand, the physics far away from the defect is governed by the bulk CFT, which is defined by an infinite set of local operators $\Om$ that satisfy unitarity and crossing symmetry.
As usual, we unit-normalize two-point functions such that the normalization of the three-point function contains dynamical information:\footnote{In this work, only symmetric-traceless operators play a role, and it is convenient to employ index-free notation $\Om(x,\eta) = \Om(x)^{\mu_1\ldots\mu_\ell} \eta_{\mu_1} \ldots \eta_{\mu_\ell}$ where $\eta \cdot \eta = 0$. Below we also introduce $\vec w$ with $\vec w^2 = 0$ to denote polarization under transverse rotations.}
\begin{align}
\label{eq:bulk-norm}
 \langle \Om_1(x_1, \eta_1) \Om_2(x_2, \eta_2) \rangle
 & = \frac{\delta_{\Om_1,\Om_2} }{x_{12}^{2\Delta_1}} \left(
        \eta_1 \cdot \eta_2
        - \frac{2 \, \eta_1 \cdot x_{12} \, \eta_2 \cdot x_{12}}
               {x_{12}^2}
    \right)^\ell \, , \\
 \langle \Om_1(x_1) \Om_2(x_2) \Om_3(x_3, \eta) \rangle
 & = \frac{\lambda_{\Om_1 \Om_2 \Om_3}}{
    x_{12}^{\Delta_1 + \Delta_2 - \Delta_3 + \ell}
    x_{13}^{\Delta_3 + \Delta_{12} - \ell}
    x_{23}^{\Delta_3 - \Delta_{12} - \ell}
 } \left( \frac{\eta \cdot x_{13}}{x_{13}^2}
        - \frac{\eta \cdot x_{23}}{x_{23}^2} \right)^\ell \, .
  \label{eq:3pt-norm}
\end{align}
Here and in what follows $\Delta_{12} = \Delta_1 - \Delta_2$.
The CFT data $\{ \Delta_i, \lambda_{ijk} \}$ fully determines the bulk theory.
This data has to do with local (short-distance) physics, and it is not modified by the presence of a defect.

On the other hand, there is a CFT that lives in the $p$-dimensional worldvolume of the defect.
The defect theory is consistent on its own, and in particular it is defined by a set of defect local operators $\wh{\Om}$ and the corresponding CFT data $\{ \wh \Delta_i, \wh \lambda_{ijk} \}$, which is constrained by unitarity and crossing.
Defect operators are labeled by $(\wh\Delta,j,s)$, the quantum numbers of the symmetry group preserved by the conformal defect $SO(p+1,1) \oplus SO(q)$.
To be more precise, $\wh \Delta$, $j$ are the $p$-dimensional scaling dimension and parallel spin, whereas $s$ is the transverse spin under $SO(q)$ rotations.

The structure of defect CFT is particularly interesting when the bulk and defect interact.
In this case, the defect theory does not admit a local stress tensor, because energy can be exchanged between defect and bulk.
The role of the stress-tensor is played by a defect operator called displacement operator, while the role of a conserved current is played by the tilt operator whenever the defect breaks a continuous global symmetry.
As we discuss in more detail in the main text, the displacement and tilt have fixed dimension $\wh\Delta_D = p+1$ and $\wh\Delta_t = p$, and satisfy Ward identities that relate certain OPE coefficients \cite{Billo:2016cpy,Padayasi:2021sik}.

In the present paper, our focus is on observables that involve the coupling between the bulk and defect theories.
The simplest such observable are bulk one-point functions, which are zero in the absence of a defect, but appear due to the partially-broken conformal symmetry:
\begin{align}
\begin{split}
 \vvev{ \Om(x, \eta) }
 & = \frac{a_\Om}{|\vec x|^{\Delta}}
 \left(\frac{(\vec \eta \cdot \vec x)^2}{\vec x \cdot \vec x}
       - \vec \eta \cdot \vec \eta \right)^{\ell/2}\, .
 \label{eq:one-pt-fun}
\end{split}
\end{align}
Here and below the double-bracket notation $\vvev{ \Om }$ indicates that an expectation value is taken in the presence of a defect.
Furthermore, we split vectors $x = (\hat x, \vec x)$ into parallel coordinates $\hat x$ and orthogonal coordinates $\vec x$.
Since bulk operators are unit normalized far away from the defect \eqref{eq:bulk-norm}, the one-point function coefficient $a_\Om$ contains dynamical information about the DCFT.
Note that \eqref{eq:one-pt-fun} only makes sense for even spin $\ell$, and odd-spin operators have vanishing one-point functions.\footnote{An important exception are parity-breaking defects in codimension two, see \cite{Gimenez-Grau:2021wiv} for a thorough discussion.}

More generally, given a bulk and a defect operator, their two-point function is kinematically fixed and the normalization $b_{\Om\wh\Om}$ contains dynamical information.
The simplest example is a bulk scalar coupled to an operator with transverse spin $s$:
\begin{align}
\begin{split}
 \vvev{ \Om(x_1) \wh\Om(\hat x_2, \vec w) }
 & = b_{\Om\wh\Om}
   \frac{(\vec x_1 \cdot \vec w)^s}
        {(\vec x_1^2 + \hat x_{12}^2)^{\wh\Delta}
          |\vec x_1|^{\Delta-\wh\Delta+s}}\, .
  \label{eq:def-b}
\end{split}
\end{align}
In fact, the CFT data associated to the defect $\{ \wh\Delta_i, \wh \lambda_{ijk}, a_i, b_{ij} \}$ is complete, in the sense that any correlation function can in principle be reconstructed from it using operator product expansions.

The bootstrap philosophy is somewhat reversed compared to this discussion.
The idea is to consider higher-point functions, which contain infinite amount of CFT data through the OPE, and to constrain $\{ \wh\Delta_i, \wh \lambda_{ijk}, a_i, b_{ij} \}$ demanding compatibility of the different OPE channels.
In a favorable situation, such as the one described in section \ref{sec:bootstrap}, one can completely fix unknown CFT data, typically in a perturbative regime.

There have been mostly two approaches in analytic studies of defects.
The first, which we do not pursue here, focuses mostly on the theory that lives on the defect, see for example work on monodromy defects \cite{Gaiotto:2013nva,Giombi:2021uae,Gimenez-Grau:2022czc} or holographic defects at strong coupling \cite{Liendo:2018ukf,Gimenez-Grau:2019hez,Bianchi:2020hsz,Ferrero:2021bsb,Cavaglia:2021bnz,Cavaglia:2022qpg}.
From the bootstrap perspective, one can consider crossing of defect four-point functions, which enjoys positivity but unfortunately it is hard to leverage known information from the bulk.
Alternatively, our approach is to consider two-point functions of bulk operators in the presence of the defect, see for example \cite{Liendo:2012hy,Soderberg:2017oaa,Bissi:2018mcq,Gimenez-Grau:2020jvf,Barrat:2020vch,Gimenez-Grau:2021wiv,Barrat:2021yvp}.
The two-point crossing equation does not have positivity, but on the flip side it allows one to extract information from the bulk to understand the defect dynamics.
The two-point functions are useful observables, because by means of operator product expansions, they capture dynamical information of infinite families of operators in the bulk and in the defect.

\subsection{Bulk two-point functions}
\label{sec:block-exp-2pt}

In the present paper we focus on the simplest defect setup that allows for the bootstrap program.
These are two-point function of bulk scalar operators\footnote{Another interesting possibility is to consider external operators that are not scalars, see for instance \cite{Lauria:2018klo,Herzog:2020bqw}.}
\begin{align}
 \vvev{ \Om_1(x_1) \Om_2(x_2) }
 = \frac{ F(r,w) }{|\vec x_1|^{\Delta_1} |\vec x_2|^{\Delta_2}} \, .
 \label{eq:two-pt-def}
\end{align}
Here we use the cross-ratios $r$, $w$ introduced in \cite{Lauria:2017wav}, but with the notation of \cite{Lemos:2017vnx}:
\begin{align}
 r + \frac{1}{r}
 = \frac{\hat x_{12}^2 + \vec x_1^{\,2} + \vec x_2^{\,2}}
        {|\vec x_1| |\vec x_2|} \, , \qquad
 w + \frac{1}{w}
 = \frac{2 \, \vec x_1 \cdot \vec x_2}{|\vec x_1| |\vec x_2|} \, .
 \label{eq:cross-ratios-rw}
\end{align}
In certain cases, it is more convenient to work with
\begin{align}
 z = rw \, , \qquad \zb = \frac{r}{w} \, ,
 \label{eq:cr-zzb}
\end{align}
which have a similar interpretation to the $z,\zb$ cross-ratios of four-point functions.
In Euclidean signature $z,\zb$ are complex coordinates on a plane orthogonal to the defect, so $r$ corresponds to the radial coordinate and $w$ to a phase.
In Lorentzian signature $z,\zb$ are real and independent.
It is also convenient to introduce the variable $\xi$:
\begin{align}
 \xi
 = \frac{(1-rw)(w-r)}{rw}
 = \frac{x_{12}^2}{|\vec x_1| |\vec x_2|}
 \, .
 \label{eq:cr-xi}
\end{align}

After these preliminary remarks, let us consider the two possible OPE expansions of the correlator $F(r,w)$.
The first possibility is to expand the bulk operator in defect modes, schematically as $\Om \sim \sum_{\wh\Om} b_{\Om\wh\Om} \wh\Om$.
For a concrete example, in the localized magnetic line defect the first terms in the defect expansion of the order parameter $\phi$ read
\begin{align}
 \lim_{|\vec x| \to 0} \phi(x)
 \sim \frac{a_\phi}{|\vec x|^{\Dp}}
 + \frac{b_{\phi\wh\phi}}{|\vec x|^{\Dp-\wh\Delta_{\wh\phi}}} \, \wh \phi
 + \frac{b_{\phi D}}{|\vec x|^{\Dp-\wh\Delta_D+1}} \, \vec x \cdot \vec D
 + \ldots \, .
 \label{eq:def-ope-phi}
\end{align}
The first term is the contribution of the defect identity $a_{\phi} = b_{\phi\wh{\mathds{1}}}$, while the third term is the contribution of the displacement operator.

As usual, the defect expansion can be organized into conformal primaries and descendants, where the contribution of descendants can be resummed.
When applied to the two-point function \eqref{eq:two-pt-def}, this leads to the defect-channel conformal block decomposition
\begin{align}
 F(r,w)
 & = \sum_{\wh \Om} 2^{-s} \, b_{\Om_1\wh\Om} \, b_{\Om_2\wh\Om}
 \, \wh f_{\Dh,s}(r,w) \, ,
 \label{eq:def-block-exp}
\end{align}
where the sum runs over defect primaries, and the blocks capture contributions from all descendants.
The somewhat awkward factor $2^{-s}$ ensures that, with the definition of $b_{\Om\wh\Om}$ in \eqref{eq:def-b}, the blocks enjoy a nice normalization \cite{Billo:2016cpy}:
\begin{align}
\label{eq:def-block}
 \wh f_{\Dh,s}(r,w)
 = r^{\wh\Delta} \, _2F_1\left(\frac p2, \wh\Delta;
  \wh\Delta+1 - \frac p2; r^2\right)
  w^{-s} \, _2F_1\left(-s, \frac{q}{2}-1;2-\frac{q}{2}-s;w^2\right) \, .
\end{align}

The correlator $F(r,w)$ admits a second expansion when the external operators approach each other.
This is the usual OPE limit $\Om_1 \Om_2 \sim \sum_{\Om} \lambda_{12\Om} \Om$, where again the expression is schematic.
When applied to the two-point function, the OPE leads to the conformal block decomposition
\begin{align}
 \xi^{\frac{\Delta_1+\Delta_2}{2}} F(r,w)
 & = \sum_{\Om, \, \ell \text{ even}} 2^{-\ell} \, \lambda_{\Om_1\Om_2\Om} \, a_{\Om} \,
     f^{\Delta_{12}}_{\Delta_\Om,\ell_\Om}(r,w) \, .
  \label{eq:blk-block-exp}
\end{align}
The sum runs over even-spin operators because odd-spin operators have vanishing one-point function, see \eqref{eq:one-pt-fun}.
Here it was natural to multiply the left-hand side by $\xi^{\frac{\Delta_1+\Delta_2}{2}}$ because then the conformal blocks only depend on $\Delta_{12}$.
Whenever the external operators are equal, we drop the extra label and use $f_{\Delta,\ell}$ to unclutter the notation.
Once again, the factor $2^{-\ell}$ is compatible with the standard definitions \eqref{eq:3pt-norm} and \eqref{eq:one-pt-fun}, and what we think is the most natural normalization for the blocks.
Although they cannot be written as simple special functions, the conformal blocks are computable from the following infinite sum \cite{Isachenkov:2018pef,Liendo:2019jpu}:
\begin{align}
 f^{\Delta_{12}}_{\Delta,\ell}(r,w)
 = & \; r^{-\frac{1}{2} \Delta_{12}}
   \sum_{m,n=0}^\infty
   \frac{4^{m-n}
    \left(\frac{-\ell}{2}\right)_m}{m! n!} \notag \\
&  \frac{
    \left(\frac{\Delta-1}{2}\right)_n
    \left(\frac{1-\ell-p}{2}\right)_m
    \left(\frac{\Delta-p}{2}\right)_n
    \left(-\frac{\ell+\Delta +2+\Delta_{12}}{2} \right)_m
    \left( \frac{\ell+\Delta-\Delta_{12}}{2} \right)_n
    \left( \frac{\ell+\Delta+\Delta_{12}}{2} \right)_{n-m} }
   {\left(-\frac{d}{2}-\ell+2\right)_m
    \left(-\frac{d}{2}+\Delta +1\right)_n
    \left(-\frac{\ell+\Delta-3}{2} \right)_m
    \left(\frac{\ell+\Delta+1}{2} \right)_n
    \left(\frac{\ell+\Delta-1}{2} \right)_{n-m}} \notag \\
& \left(1-r^2\right)^{\ell-2 m}
  {}_2F_1\left( \!\!
  \begin{array}{c c}
   \frac{\Delta +\ell-\Delta_{12}}{2}-m+n, \,
   \frac{\Delta +\ell}{2}-m+n \\
   \Delta +\ell-2 m+2 n
  \end{array}; 1-r^2 \right) \notag \\
& \left(\tfrac{(w-r)(1-r w)}{w}\right)^{\frac{\Delta-\ell}{2}+m+n}
  {}_4F_3\left( \!\! \begin{array}{c}
    -n, \, -m, \,
    \frac{\Delta_{12}+1}{2}, \,
    \frac{\Delta_{12}+\Delta -\ell-d+2}{2} \\
    \frac{\Delta_{12}-\Delta -\ell-2n+2}{2}, \,
    \frac{\Delta_{12}+\Delta +\ell-2m}{2}, \,
    \frac{\Delta -\ell-d+3}{2}
    \end{array} ; 1 \right) \, .
 \label{eq:bulk-block-ser}
\end{align}
Whenever we compute conformal block decompositions, we also find it convenient to use the lightcone representation \eqref{eq:lightcone-exp-block}.

To summarize, the two-point function $F(r,w)$ can be expanded in two different ways. When one operator approaches the defect we have $r \to 0$ and it is natural to use the defect-channel decomposition \eqref{eq:def-block-exp}.
On the other hand, when the two operators approach each other we have $\xi \to 0$ and it is natural to use the bulk-channel decomposition \eqref{eq:blk-block-exp}.
The equivalence of these two expansions can be thought of as crossing equation, that can be used to solve theories of interest.
In this work we use two analytic tools to solve crossing, the dispersion relation \eqref{eq:disp-rel}, and the inversion formula that we now discuss.

\subsection{Lorentzian inversion formula}

The Lorentzian inversion formula is a tool for reconstructing the defect CFT data, which is encapsulated in a function $b(\Dh, s)$, starting from the single discontinuity of a correlator $\Disc F$.
The crucial property, emphasized in section \ref{sec:struct-calc}, is that the discontinuity is a simpler object than $F$, specially in perturbative settings.
For the application in section \ref{sec:inv-P2}, we have found it more convenient to work in $z,\zb$ coordinates, when the inversion formula reads \cite{Lemos:2017vnx}
\begin{align}
\label{eq:inversion}
 b(\Dh, s)
 &= \int_0^1 \frac{dz}{2z} z^{-\frac{\Dh-s}{2}} \int_1^{1/z}
    \frac{d\zb}{2\pi i}
    (1 - z \bar z) (\bar z - z)
    \bar z^{-\frac{\Dh + s + 4}{2}} \\
 & \quad \times
   {}_2F_1\left(1-\Dh,1-\frac{p}{2},-\Dh+\frac{p}{2}+1,z \zb\right)
   {}_2F_1\left(s+1,2-\frac{q}{2},\frac{q}{2}+s,\frac{z}{\zb}\right)
   \text{Disc} \, F(z, \zb) \, , \notag
\end{align}
and the discontinuity is computed around the cut $\zb \in [1, \infty)$
\begin{align}
\label{eq:disc-def}
 \text{Disc} \, F(z, \zb)
 = F(z, \zb + i0) - F(z, \zb - i0), \qquad \zb \ge 1 \, .
\end{align}
Note that the discontinuities in terms of $r,w$ and $z,\zb$ are negative of each other $\Disc_\zb = -\Disc_w$, we hope this does not create confusion.

To recover the CFT data from $b(\Dh,s)$, observe that the location of the poles are scaling dimensions, while the residues are OPE coefficients:
\begin{equation}
b(\Dh,s)
= - \sum_{\wh\Om}
    \frac{2^{-s} \, b_{\Om_1\wh\Om} b_{\Om_2\wh\Om}}
         {\Dh - \wh \Delta_{\wh\Om}} \,.
\end{equation}
Since we work in pertubation theory, we can split the dimensions into a tree-level and a perturbative correction $\wh \Delta_{\wh\Om} = \wh \Delta^{\text{tree}}_{\wh\Om} + \wh \gamma_{\wh\Om}$.
In this case, the anomalous dimension comes from the double poles:
\begin{equation}
b(\Dh,s)
= - \sum_{\wh\Om} 2^{-s} \, b_{\Om_1\wh\Om} \, b_{\Om_2\wh\Om} \left(
    \frac{1}{\Dh - \wh \Delta^{\text{tree}}_{\wh\Om}}
  + \frac{\wh \gamma_{\wh\Om}}
         {\big( \Dh - \wh \Delta^{\text{tree}}_{\wh\Om} \big)^2}
  \right) \,.
 \label{eq:poles-iform}
\end{equation}
In section \ref{sec:inv-P2} we extract the defect CFT data using this formalism, and then reconstruct the correlator summing the defect expansion \eqref{eq:def-block-exp}.
As shown in \cite{Barrat:2022psm,Bianchi:2022ppi}, this is completely equivalent to applying the integral transform $\DR[F(r,w)]$.

\section{Discontinuity of \texorpdfstring{$\phi^2$}{phi2}}
\label{sec:disc}

The purpose of this appendix is to compute the discontinuity of the bulk-channel conformal block corresponding to $\phi^2 \equiv \phi_a \phi_a$.
We find it more convenient to use $z = rw$ and $\zb =r/w$ coordinates.
Let us compute the discontinuity for a scalar block $\Delta=2\Dp+\gamma$, where for simplicity we take the external operators to be equal.
From the series representations of bulk blocks \eqref{eq:bulk-block-ser} or \eqref{eq:lightcone-exp-block}, we observe that removing the prefactor
\begin{align}
\label{eq:red-blocks}
 f_{\Delta,0}(z,\zb)
 = \big[(1-z) (1-\zb)\big]^{\frac{\Delta}{2}}
   \tilde f_{\Delta,0}(z, \zb) \, ,
\end{align}
the expansion of $\tilde f_{\Delta,0}(z, \zb)$ is in integer powers of $(1-z),(1-\zb)$.
With this information, we compute the discontinuity assuming $\gamma$ is small:
\begin{align}
  \Disc \xi^{-\frac12\Dp} f_{\Delta,0}(z,\zb)
 & = (z \zb)^{\frac{1}{2}\Dp}
     \tilde f_{\Delta,0}(z,\zb)
     \Disc \! \big[ (1-z)(1-\zb) \big]^{\frac12 \gamma} \notag \\
 & \approx
     \frac12 \gamma (z \zb)^{\frac{1}{2}\Dp}
     \tilde f_{\Delta,0}(z,\zb)
     \Disc \log (1-\zb) \notag \\
 & = - i \pi \gamma (z \zb)^{\frac{1}{2}\Dp}
     \tilde f_{\Delta,0}(z,\zb) \, .
 \label{eq:disc-calc}
\end{align}
Remember that the discontinuity is computed around the cut $\zb \in [1, +\infty)$, see \eqref{eq:disc-def}.
In the second line we Taylor expanded to leading order in $\gamma$, which is the only term we need in this paper.
With similar manipulations, one can prove \eqref{eq:disc-block}.

Let us now compute the discontinuity of $\phi^2$.
Note that the discontinuity \eqref{eq:disc-calc} is order $\veps$ because $\gamma \propto \veps$.
We can evaluate the result for $\veps=0$, which simplifies the calculation, and all is left is to obtain $\tilde f_{2,0}$ in $d=4$ dimensions.
Using the series representation \eqref{eq:bulk-block-ser} we find
\begin{align}
\label{eq:scalar-block-exp}
 \tilde f_{2,0}(z,\zb)
 & = \sum_{n=0}^\infty
 \frac{\big[(1-z) (1-\zb) \big]^{n}}{4^n(2n+1)} \,
 {}_2F_1\left(n+1,n+1;2 n+2;1-z \zb\right) \, .
\end{align}
It is hard to compute this sum in full generality, but fortunately in section \ref{sec:inv-P2} we only need its expansion around $z = 0$ to all orders in $\zb$.
This can be achieved by expanding the summand in \eqref{eq:scalar-block-exp} as $z \to 0$ and then summing over $\zb$.
For simplicity, let us first look at the leading-order term $z^0$
\begin{align}
 \tilde f_{2, 0}(z,\zb) \big|_{z^0}
&= -\sum_{n=0}^\infty \frac{\Gamma(n+\frac{1}{2})}{\sqrt{\pi } \Gamma (n+1)}
   (1-\zb)^n \big(2 H_n + \log z\zb \big) \notag \\
&= \frac{1}{\sqrt{\zb}} \log \left(\frac{16\zb^2}{(1 + \sqrt{\zb})^4}\right)
 - \frac{1}{\sqrt{\zb}} \log z\zb \, .
 \label{eq:easy-sum}
\end{align}
One can similarly expand \eqref{eq:scalar-block-exp} to higher orders in $z$ and compute the $\zb$ sums order by order.
A useful trick is to notice that all sums are essentially of the form \eqref{eq:easy-sum} after acting with certain $\zb$ derivatives.
This observation allows us to compute the expansion to high orders, and here we simply show the first few terms:
\begin{align}
\begin{split}
 \tilde f_{2,0}(z, \zb)
 & = \frac{1}{\sqrt{\zb}} \log \left( \frac{16 \zb }{z(1 + \sqrt{\zb})^4}\right)
    \left(
        1
        + z \frac{(\zb+1)^2}{4 \zb}
        + z^2 \frac{\left(3 \zb^2+2 \zb+3\right)^2}{64 \zb^{2}}
        + \ldots
    \right) \\
 & \quad
   - \frac{\left(\sqrt{\zb}-1\right)^2}{2 \zb^{3/2}} \left(
      z (\zb+1)
    + z^2 \frac{\left(7 \zb+2 \sqrt{\zb}+7\right) \left(3 \zb^2+2 \zb+3\right)}{32 \zb}
    + \ldots
 \right) \, .
\end{split}
\end{align}
Restoring appropriate prefactors, this gives \eqref{eq:disc-full}.

\section{Block expansions}
\label{sec:block-exp}

In this appendix we present the conformal block decomposition of the two-point functions \eqref{eq:PP-diag}, \eqref{eq:PP2-diag} and \eqref{eq:P2P2-diag}.
Here we consider each of the building blocks separately, and we present the combined CFT data in section \ref{sec:ope-analysis}.

To obtain the conformal block decomposition it is convenient to work with $z = rw$ and $\zb = r/w$, because the expansion is $z,\zb$ naturally organizes by the twist.
For example, in the bulk channel we consider the lightcone limit $|1-z| \ll |1-\zb| \ll 1$, and leading order in $1-z$ and all orders in $1-\zb$ gives the CFT data of the leading-twist family.
To obtain the CFT data, we expand the correlator and the conformal blocks to high order in $1-\zb$ using \mathematica.
Solving for the unknown coefficients, it is often possible to guess a closed form expression.

\subsection{Bulk channel}

\subsubsection{Lightcone blocks}

We compute the bulk-channel expansions in the limit $|1-z| \ll |1-\zb| \ll 1$.
Since we organize our calculations twist by twist, it is convenient expand the conformal blocks as follows
\begin{align}
 f^{\Delta_{12}}_{\Delta,\ell}(z,\zb)
 = \sum_{n=0}^\infty \sum_{q=-n}^n
   A_{n,q}(\Delta,\ell) \,
   (1-z)^{\frac{\Delta-\ell}{2} + n} \,
   \tilde k_{\frac{\Delta+\ell}{2} + q}^{\Delta_{12}}(1-\zb) \, .
 \label{eq:lightcone-exp-block}
\end{align}
Similar expansions have been used in the literature, for example \cite{Simmons-Duffin:2016wlq,Caron-Huot:2017vep}, but our chiral blocks take a somewhat unusual form:
\begin{align}
 \tilde k_{\bar h}^{\Delta_{12}}(1-\zb)
 = (1-\zb)^{\bar h} \, \zb^{\Delta_{12}/4} \, _2F_1\left(\bar h,\bar h+\frac{\Delta_{12}}{2};2 \bar h;1-\zb\right) \, .
\end{align}
The coefficients $A_{n,q}(\Delta,\ell)$ in the lightcone expansion could be fixed with \eqref{eq:bulk-block-ser}, but in practice we found even more convenient to fix them directly from the Casimir equation \cite{Billo:2016cpy}
\begin{align}
 \Bigg[ & (1-z)^2 \left(
    z \partial_z^2
    + \partial_z
    -\frac{\Delta_{12}^2}{16 z}
\right) \notag \\
& -\frac{z (1-z) (1-\zb)}{z-\zb}
\left( p \frac{(1-z) (1+\zb)}{1-z \zb} - (d-2) \right)
\partial_z
+ (z \leftrightarrow \zb)
\Bigg] f_{\Delta, \ell}
= \frac{c_{\Delta,\ell}}{2} f_{\Delta, \ell} \, ,
\end{align}
where $c_{\Delta,\ell} = \Delta(\Delta-d) + \ell(\ell+d-2)$.
We are mostly interested in the leading-twist family of operators, for which only $A_{0,0}(\Delta,\ell) = 1$ is needed.
When we need the first subleading trajectory, the expansion follows from
\begin{align}
 A_{1,-1}(\Delta,\ell)
 & = \frac{\ell (d-2 p-2)}{2 \ell + d-4} \, , \\
 A_{1,0}(\Delta,\ell)
 & = \frac{\Delta -\ell}{4} \, , \\
 A_{1,1}(\Delta,\ell)
 & = \frac{(\Delta -1) (d-2 p-2) (\Delta -\Delta_{12}+\ell)
           (\Delta +\Delta_{12}+\ell)}
          {16 (2 \Delta +2-d) (\Delta +\ell-1) (\Delta +\ell+1)} \, .
\end{align}
The coefficients $A_{n\ge2,q}$ can also be generated efficiently with the help of a computer.

\subsubsection{Correlator \texorpdfstring{$\vvev{ \phi \phi }$}{ <phi phi>}}

First we should decompose the correlator $F_{\phi\phi}(r,w)$ in the two symmetry channels \eqref{eq:blk-tens-struct}, and then we should multiply by $\xi^\Dp$ as in equation \eqref{eq:blk-block-exp}.
The simplest piece just gives a constant $\xi^\Dp \times \xi^{-\Dp} = 1$, namely it is the conformal block of the bulk identity.
The term $H(r,w)$ has been discussed in \eqref{eq:exp-H-PP}, and only corrects the scalar of dimension $\Delta=2$, namely $\phi^2$ or $T_{ab}$.
Finally, the constant piece generates the expansion
\begin{align}
  \xi^\Dp
& =
    \sum_{\substack{\ell=0 \\ \ell \text{ even}}}^\infty
    \frac{\big(\frac{1}{2} \Dp\big)_{\ell/2}^2}{
          2^\ell \, \frac{\ell}{2}! \,
          \big(\frac{2\Dp+\ell-1}{2} \big)_{\ell/2}}
    \Bigg(
        f_{2\Dp+\ell,\ell}
      + \frac{(\ell+1)^2 \veps}{8 (\ell+2) (2 \ell+1)}
        f_{2\Dp+\ell+2,\ell}
      + O(\veps^2)
    \Bigg)
    \, .
 \label{eq:exp-bid-PP}
\end{align}
In this expansion the blocks are evaluated in $d=4-\veps$ dimensions, and we have found that no higher-twist families besides $n=0,1$ are involved.

\subsubsection{Correlator \texorpdfstring{$\vvev{ \phi \phi^2 }$}{ <phi phi2>}}

For the $F_{\phi\phi^2}$ correlator we should multiply by $\xi^{\frac{1}{2}(\Dp+\Dpp)}$ before expanding, and remember to use blocks with unequal external dimensions.
For example, the constant term in the correlator gives
\begin{align}
  \xi^{\frac{1}{2}(\Dp+\Delta_{\phi^2})}
& =
    \sum_{\substack{\ell=0 \\ \ell \text{ even}}}^\infty
    \frac{\left(\frac{1}{2} \Dp\right)_{\ell/2}
          \left(\frac{1}{2} \Delta_{\phi^2} \right)_{\ell/2}}
         {2^{\ell} \, \frac{\ell}{2}!
          \left(\frac{\Dp+\Delta_{\phi^2}+\ell-1}{2}\right)_{\ell/2}}
    \bigg(
        f^{\Delta_\phi-\Delta_{\phi^2}}_{\Delta_\phi+\Delta_{\phi^2}+\ell,\ell}
      + \frac{\veps}{16}
        f^{\Delta_\phi-\Delta_{\phi^2}}_{\Delta_\phi+\Delta_{\phi^2}+\ell+2,\ell} \notag \\
& \qquad \qquad \qquad \qquad \qquad
      + \frac{(\ell+2) (\ell+6) \veps}{12288 (\ell+3) (\ell+5)}
        f^{\Delta_\phi-\Delta_{\phi^2}}_{\Delta_\phi+\Delta_{\phi^2}+\ell+6,\ell}
      + \ldots
    \bigg)
    \, .
 \label{eq:exp-bid-PP2}
\end{align}
Note that $\ldots$ stands for the higher-twist families that we omit, as well as corrections at $O(\veps^2)$.

Next we consider the function $H(r,w)$. The expansion of $H(r,w)$ follows from the combination of \eqref{eq:exp-H-PP} with \eqref{eq:scalar-block-exp}, but now it must match an expansion of blocks with unequal external dimensions.
Because $H(r,w)$ is multiplied by $\veps$, it is sufficient at the order we work to consider $d = 4$ and tree-level dimensions $\Dp-\Dpp=-1$.
The result only involves the twist-three family
\begin{align}
 \xi^{3/2} \, H(r,w)
 = \sum_{\substack{\ell=0 \\ \ell \text{ even}}}^\infty
   \frac{4^{-\ell}}{\ell+1} \left(
   \frac12 H_{\frac{\ell}{2}} - \frac12 H_{\frac{\ell+1}{2}} - \log 4
 + \partial_\Delta \right) f^{-1}_{3+\ell,\ell} \, .
 \label{eq:exp-H-PP2}
\end{align}

To expand the remaining terms, remember that by multiplet recombination $\phi^3$ becomes a descendant of $\phi$. As a result, the $\veps$--expansion of $f_{\Dp,0}^{\Dp-\Dpp}$ decomposes into blocks $f_{1,0}^{-1}$, $f_{3,0}^{-1}$ and their derivatives.
This decomposition turns out to be very subtle.
In particular, we observed that it is sensitive to the values of scaling dimensions up to three-loops \cite{Kleinert:1991rg}:
\begin{align}
 \Dp
&= 1
 -\frac{\veps}{2}
 + \frac{(N+2) \veps^2}{4 (N+8)^2} \left(
   1 - \frac{N^2-56 N-272}{4 (N+8)^2} \, \veps
 + O(\veps^2)
 \right)
 \, , \\
 \Dpp
&= 2
 - \veps
 + \frac{(N+2) \veps}{N+8} \left(
   1
 + \frac{13 N+44}{2 (N+8)^2} \, \veps
 + O(\veps^2)
 \right) \, .
\end{align}
The lightcone expansion \eqref{eq:lightcone-exp-block} of $f_{\Dp,0}^{\Dp-\Dpp}$ can be compared with an ansatz of $f_{1,0}^{-1}$ and $f_{3,0}^{-1}$ blocks.
To simplify later calculations, it is useful to organize the result as
\begin{align}
 f_{\Dp,0}^{\Dp-\Dpp}(r,w)
&= \xi^{\frac12\Dp}
 - \frac{N+2}{2 (N+8)} \, \veps \, \xi^{1/2} \log \frac{4 r}{(r+1)^2}
 - \frac{3+3 \log 2}{4} \, \veps \, f_{3,0}^{-1}
 + \frac{3\veps}{4} \partial_\Delta f_{3,0}^{-1} \notag \\
& \quad
 + \left(\frac{N+8}{8} + \frac{N^2-3 N-22}{16 (N+8)} \, \veps
    +\frac{3 \log 2}{4} \, \veps \right) f_{\Dp+\Dpp,0}^{\Dp-\Dpp}
 + O(\veps^2) \, .
 \label{eq:exp-Pblock}
\end{align}
Note that the first two terms also correspond to blocks of the aforementioned form
\begin{align}
 \xi^{\frac12\Dp}
 = f_{\Dp,0}^{-\Dp} \, , \qquad
 \xi^{1/2} \, \log \frac{4 r}{(1+r)^2}
&= 2 \partial_{\Delta_{12}} f_{1,0}^{-1}
 - \frac18 f_{3,0}^{-1} \, ,
 \label{eq:exp-log-PP2}
\end{align}
but in practice it is easier to use them in the form \eqref{eq:exp-Pblock}.

\subsubsection{Correlator \texorpdfstring{$\vvev{ \phi^2 \phi^2 }$}{ <phi2 phi2>}}
\label{sec:app-blk-P2P2}

This correlator should be multiplied by $\xi^\Dpp$ before expanding, so in particular the constant piece gives
\begin{align}
  \xi^{\Delta_{\phi^2}}
& = \sum_{\substack{\ell=0 \\ \ell \text{ even}}}^\infty
    \frac{\left(\frac{1}{2} \Delta_{\phi^2} \right)_{\frac{\ell}{2}}^2}{
          2^\ell \, \frac{\ell}{2}!
          \left(\frac{2\Delta_{\phi^2} + \ell-1}{2} \right)_{\frac{\ell}{2}}}
    \left(
        f_{2\Delta_{\phi^2}+\ell,\ell}
      + \frac{(\ell+2) \veps}{8 (2 \ell+3)}
        f_{2\Delta_{\phi^2}+\ell+2,\ell}
      + \ldots
    \right) \, ,
 \label{eq:exp-bid-P2P2}
\end{align}
where $\ldots$ stand for higher-twist families as well as higher orders in $\veps$.
Next there is a term $\xi^{\Dp-\Dpp}$, but its expansion has already been presented in \eqref{eq:exp-bid-PP}.
Out of the two terms involving $H(r,w)$, one is expanded using \eqref{eq:exp-H-PP}, while for the other we find
\begin{align}
  \xi^2 \, H(r,w)
& = \sum_{n=0}^\infty
    \sum_{\substack{\ell=0 \\ \ell \text{ even}}}^\infty
    \big({-} A_{\ell,n} + B_{\ell,n} \partial_\Delta \big)
    f_{4+\ell+2n,\ell} \, .
 \label{eq:exp-xi-P2P2}
\end{align}
Again we are working at order $O(\veps^0)$ because this term is multiplied by $\veps$ in the correlator.
In this case, we have not found closed formulas for the coefficients $A_{\ell,n}$ and $B_{\ell,n}$, although it is possible to generate them using certain recursion relations.
Since these coefficients do not play an important role in the main text, we content ourselves by presenting the lowest-lying data in tables \ref{tab:1} and \ref{tab:2}.
We observed empirically that for $n$ odd the coefficients vanish.

\begin{table}
\renewcommand{\arraystretch}{1.1}
 \begin{align*}
\begin{array}{c|l|l}
 \ell & A_{\ell,0} & A_{\ell,2} \\ \hline
 0 & 1+\log 2 & \frac{11}{10368}+\frac{\log 2}{864} \\
 2 & \frac{101}{1800}+\frac{7 \log 2}{120} & \frac{2473}{34992000}+\frac{31 \log 2}{388800} \\
 4 & \frac{119521}{38102400}+\frac{407 \log 2}{120960} & \frac{50109481}{11491749068800}+\frac{1437 \log 2}{287006720} \\
 6 & \frac{124370867}{692583091200}+\frac{3023 \log 2}{15375360} & \frac{7111918573}{26797814184960000}+\frac{59303 \log 2}{192471552000} \\
 8 & \frac{1004193021703}{96075126411264000}+\frac{1456787 \log 2}{125462937600} & \frac{7040698918075}{436903886309069684736}+\frac{9199735 \log 2}{487966670585856} \\
 10 & \frac{143434115533099}{233070570663613562880}+\frac{11072177 \log 2}{16019107872768} & \frac{14795257077824297}{15110820472963896705024000}+\frac{5202131 \log 2}{4515554582855680}
\end{array} \\[-3em]
\end{align*}
\caption{Low-lying coefficients defined by equation \eqref{eq:exp-xi-P2P2}.}
\label{tab:1}
\end{table}

\begin{table}
\renewcommand{\arraystretch}{1.3}
 \begin{align*}
\begin{array}{c|cccccc}
 \ell & 0 & 2 & 4 & 6 & 8 & 10 \\ \hline
 B_{\ell,0} & 1 & \frac{7}{120} & \frac{407}{120960} & \frac{3023}{15375360} & \frac{1456787}{125462937600} & \frac{11072177}{16019107872768} \\ \hline
 B_{\ell,2} & \frac{1}{864} & \frac{31}{388800} & \frac{1437}{287006720} & \frac{59303}{192471552000} & \frac{9199735}{487966670585856} & \frac{5202131}{4515554582855680}
\end{array} \\[-3em]
\end{align*}
\caption{Low-lying coefficients defined by equation \eqref{eq:exp-xi-P2P2}.}
\label{tab:2}
\end{table}

Finally, the expansion of the logarithmic term at $\veps = 0$ contains only twist-two and twist-four operators:
\begin{align}
  \xi \, \log \frac{4 r}{(1+r)^2}
  = - \sum_{\substack{\ell=0 \\ \ell \text{ even}}}^\infty
   \frac{(\ell!)^2}
        {8^\ell (\frac{\ell}{2}!)^3 \left(\frac{\ell+1}{2}\right)_{\frac{\ell}{2}}} \left(
    H_\ell \, f_{2+\ell,\ell}
  + \frac{(\ell+1)^2}{4 (\ell+2) (2 \ell+1)} f_{4+\ell,\ell}
\right) \, .
\label{eq:exp-log-P2P2}
\end{align}

\subsection{Defect channel}
\label{sec:app-def-exp}

We now expand the same terms but in the defect channel with a similar strategy as above.
The only difference is that we consider the limit $|z| \ll |\zb| \ll 1$, and the defect expansion is organized by transverse twist $\Dh-s$.
The defect CFT data is somewhat easier to extract than the bulk one, since we have closed form expressions for the blocks \eqref{eq:def-block}.
Note that we should no longer multiply the correlator by $\xi^{\frac{\Delta_1+\Delta_2}{2}}$ before expanding.

\subsubsection{Correlator \texorpdfstring{$\vvev{ \phi \phi }$}{ <phi phi>}}

The constant term corresponds to the exchange of the defect identity $\wh f_{0,0} = 1$, while the expansion of $H(r,w)$ appears in \eqref{eq:expd-H}.
The only term left is the bulk identity, which is well-known in the literature \cite{Lemos:2017vnx}
\begin{align}
 \xi^{-\Dp} =
 \sum_{n,s=0}^\infty
 \frac{\left(\Dp-\frac{d-2}{2}\right)_n (\Dp)_{2 n+s}}{s! n!
 \left(s + \frac{d-1}{2}\right)_{n} \left(\Dp+s+n-\frac{1}{2}\right)_n}
 \wh f_{\Dp+s+2n,s} \, .
 \label{eq:expd-bid}
\end{align}

\subsubsection{Correlator \texorpdfstring{$\vvev{ \phi \phi^2 }$}{ <phi phi2>}}

The first three terms of $F_{\phi\phi^2}$ can be treated identically as for $F_{\phi\phi}$, except that we should replace $\Dp \to \frac12\Dpp$ in \eqref{eq:expd-bid}.
The expansion of the new logarithmic term is
\begin{align}
  \xi^{-1} \, \log \frac{4 r}{(1+r)^2}
& = \sum_{s=0}^\infty \big(2 \log 2 + \partial_{\wh\Delta} \big) \wh f_{1+s,s}
  + \sum_{n=1}^\infty \sum_{s=0}^\infty
    \frac{4 (-1)^n (s+1)_n}{n (2 s+n+1) \left(\frac{2 s+3}{2}\right)_{n-1}}
    \wh f_{1+s+n,s} \, .
 \label{eq:expd-log}
\end{align}
Again have set $\veps = 0$ because the logarithmic term appears at order $O(\veps)$ in the correlator.

\subsubsection{Correlator \texorpdfstring{$\vvev{ \phi^2 \phi^2 }$}{ <phi2 phi2>}}

Finally, for the $F_{\phi^2\phi^2}$ correlator the only missing ingredient is the expansion
\begin{align}
  \xi^{-1} \, H(r,w)
& = \sum_{n=0}^\infty \sum_{s=0}^\infty \left(
    - \wh A_{s,n}
    + \wh B_{s,n} \partial_{\wh\Delta}
  \right) \wh f_{2+s+2n,s} \, .
  \label{eq:expd-xiH}
\end{align}
The coefficients $\wh B_{s,n}$ are amenable to being guessed, and for the first two we find
\begin{align}
 \wh B_{s,0}
 = H_{s+\frac{1}{2}} + \log 4 \, , \quad
 \wh B_{s,1}
 = -\frac{4 (s+1)^2}{(2 s+3)^2}
 + \frac{2 \left(2 s^2+8 s+7\right)}{(2 s+3) (2 s+5)}
 \left(H_{s+\frac12} + \log 4\right) \, .
\end{align}
On the other hand, we could not guess $\wh A_{s,n}$, so instead we present some low-lying values in table \ref{tab:3}.

\begin{table}
\renewcommand{\arraystretch}{1.1}
\begin{align*}
\begin{array}{c|l|l}
 s & \wh A_{s,0} & \wh A_{s,1} \\ \hline
 0 & 4-4 \log 2 & \frac{1624}{675}-\frac{128 \log 2}{45} \\
 1 & \frac{46}{9}-\frac{16 \log 2}{3} & \frac{177328}{55125}-\frac{2048 \log 2}{525} \\
 2 & \frac{1291}{225}-\frac{92 \log 2}{15} & \frac{7733504}{2083725}-\frac{30208 \log 2}{6615} \\
 3 & \frac{68044}{11025}-\frac{704 \log 2}{105} & \frac{5711504}{1403325}-\frac{2048 \log 2}{405} \\
 4 & \frac{1290557}{198450}-\frac{2252 \log 2}{315} & \frac{97142374696}{22319572275}-\frac{2698624 \log 2}{495495} \\
 5 & \frac{81294376}{12006225}-\frac{26032 \log 2}{3465} & \frac{1814386766048}{395665144875}-\frac{50683904 \log 2}{8783775} \\
 6 & \frac{28389178703}{4058104050}-\frac{352276 \log 2}{45045} & \frac{42085755740992}{8795940528375}-\frac{69466112 \log 2}{11486475} \\
 7 & \frac{14587073944}{2029052025}-\frac{364288 \log 2}{45045} & \frac{2549071235257504}{514101781167975}-\frac{222273536 \log 2}{35334585} \\
 8 & \frac{17260962711589}{2345584140900}-\frac{6373076 \log 2}{765765} & \frac{431808003481906928}{84463898517773775}-\frac{37772670592 \log 2}{5805264465} \\
 9 & \frac{1589834722001101}{211688968716225}-\frac{124151504 \log 2}{14549535} & \frac{12348057094959656656}{2351652753468543525}-\frac{47092860928 \log 2}{7027425405} \\
 10 & \frac{6474883235224849}{846755874864900}-\frac{126922844 \log 2}{14549535} & \frac{247285672035572912864}{45993208613755528125}-\frac{189076662784 \log 2}{27488228625} \\
\end{array} \\[-3em]
\end{align*}
\caption{Low-lying coefficients defined by equation \eqref{eq:expd-xiH}.}
\label{tab:3}
\end{table}

\providecommand{\href}[2]{#2}\begingroup\raggedright\endgroup

\end{document}